\documentclass{article}
\usepackage[colorlinks=true,linkcolor=blue,citecolor=green,urlcolor=blue]{hyperref}
\usepackage{amsmath}
\usepackage{PRIMEarxiv}
\usepackage{multirow} 
\usepackage[utf8]{inputenc} % allow utf-8 input
\usepackage[T1]{fontenc}    % use 8-bit T1 fonts
\usepackage{hyperref}       % hyperlinks
\usepackage{url}            % simple URL typesetting
\usepackage{booktabs}       % professional-quality tables
\usepackage{amsfonts}       % blackboard math symbols
\usepackage{nicefrac}       % compact symbols for 1/2, etc.
\usepackage{microtype}      % microtypography
\usepackage{lipsum}
\usepackage{fancyhdr}       % header
\usepackage{graphicx}       % graphics
\graphicspath{{media/}}     % organize your images and other figures under media/ folder
\usepackage{algorithmic}
\usepackage{algorithm}
\usepackage{pifont}  
\usepackage{color}  
\title{MedMamba: Vision Mamba for Medical Image Classification
}

\author{
  Yubiao Yue \\
  School of Biomedical Engineering\\
  Guangzhou Medical University \\
  \texttt{jiche2020@126.com} \\
   \And
  Zhenzhang Li \\
  School of Mathematics and Systems Science\\
  Guangdong Polytechnic Normal University \\
  \texttt{zhenzhangli@gpnu.edu.cn} \\
}

\begin{document}
\maketitle

\begin{abstract}
Since the era of deep learning, convolutional neural networks (CNNs) and vision transformers (ViTs) have been extensively studied and widely used in medical image classification tasks. Unfortunately, CNN's limitations in modeling long-range dependencies result in poor classification performances. In contrast, ViTs are hampered by the quadratic computational complexity of their self-attention mechanism, making them difficult to deploy in real-world settings with limited computational resources. Recent studies have shown that state space models (SSMs) represented by Mamba can effectively model long-range dependencies while maintaining linear computational complexity. Inspired by it, we proposed MedMamba, the first Vision Mamba for generalized medical image classification. Concretely, we introduced a novel hybrid basic block named SS-Conv-SSM, which purely integrates the convolutional layers for extracting local features with the abilities of SSM to capture long-range dependencies, aiming to model medical images from different image modalities efficiently. By employing the grouped convolution strategy and channel-shuffle operation, MedMamba successfully provides fewer model parameters and a lower computational burden for efficient applications without sacrificing accuracy. We thoroughly evaluated MedMamba using 16 datasets containing ten imaging modalities and 411,007 images. Experimental results show that MedMamba demonstrates competitive performance on most tasks compared with the state-of-the-art methods. This work aims to explore the potential of Vision Mamba and establish a new baseline for medical image classification, thereby providing valuable insights for developing more powerful Mamba-based artificial intelligence algorithms and applications in medicine. The source codes and all pre-trained weights of MedMamba are available at \textbf{https://github.com/YubiaoYue/MedMamba}.
\end{abstract}

% keywords can be removed

\section{Introduction}
Medical image classification is a basic step in medical image analysis and has been an essential task in computer-aided diagnosis (CAD) \cite{1}. It aims to distinguish medical images according to certain criteria (such as clinical pathology or imaging patterns) and plays a vital role in the field of healthcare, including clinical diagnosis, disease treatment, and disease monitoring \cite{2,3,4,5,6}. Benefitting from the rapid development of digital medical imaging technology, a large number of medical images from different imaging modalities, including computed tomography (CT), ultrasound (US), X-ray, microscope, endoscope, and magnetic resonance imaging (MRI), have been widely used, greatly facilitating the development of clinical medicine \cite{7}. However, the tremendous increase in medical images poses a challenge for manual classification and interpretation, as this process is very time-consuming and labor-intensive \cite{8,9}. Particularly, the ability of clinicians (such as sonographers, radiologists, and pathologists) to accurately distinguish images of various organs and lesions largely depends on their domain knowledge and clinical experience. This reliance leads to significant differences in the methods adopted and conclusions drawn by different individuals when analyzing and interpreting medical images \cite{10}.

To address this clinical challenge, many Artificial Intelligence-based CAD solutions have been developed and tested to improve the accuracy and efficiency of disease diagnosis and management \cite{11,12}. Over the past few decades, the technology used in CAD has undergone a transformation from traditional machine learning to deep learning. Due to the explosive growth in the number of medical images and the significant improvement in computing device performance, deep learning models has achieved tremendous success and shown great promise in medical image processing tasks. Nowadays, how to employ deep learning technology to effectively perform various medical image classification tasks has become a fundamental and significant task in CAD and computer vision \cite{13,14,15}. Importantly, an excellent medical image classification model can also serve as a generic vision backbone to effectively extract representative features from various advanced tasks, such as medical image segmentation, medical image object detection, and medical image reconstruction \cite{16}.

The two most popular architectures in deep learning, namely convolutional neural networks (CNNs) and vision transformers (ViTs), are dominating the field of visual representation learning and have been widely used in various image classification tasks \cite{17,18,19}. It is worth noting that compared with natural image datasets (e.g., ImageNet \cite{20} ), the intra-class variability and inter-class similarity pose an even greater challenge for medical image classification \cite{21}. Although CNNs can effectively extract local features, they have difficulty in capturing global context and long-range dependencies, leading to insufficient feature extraction and unsatisfactory classification results, as they are inherently limited by their local receptive field. ViTs, originally designed for natural language processing, have been successfully applied to computer vision tasks. In essence, ViTs are not able to handle spatial image hierarchies but represent an input image as a sequence of image patches \cite{22}. By adopting cascaded self-attention modules, vision transformers can effectively capture long-range dependencies, but unfortunately, this degrades local feature details \cite{23,24}. Recent studies have shown that combining local features in an image with corresponding long-range dependencies is the key to achieving accurate medical image classification results \cite{25,26,27,28,29}. Given this complementarity between CNNs and ViTs and the characteristics of medical images, researchers have begun to utilize hybrid architectures based on CNN and ViT (CNN-ViT) to more efficiently analyze various medical images \cite{30,31,32,33,34}. The only drawback is that the self-attention mechanism of ViT has high quadratic complexity in long sequence modeling, resulting in a high computational burden (especially for medical images with high resolution) and making it difficult to deploy the model in clinical settings with limited computing resources. Some previous works have attempted to use CNN, ViT and CNN-ViT to build a generic medical image classification model \cite{9,15,16,21,35,36}. However, these models have more or less the above limitations. Given the importance of medical image classification tasks and the limitations of existing models, exploring a generic medical image classification architecture that can achieve a good trade-off between visual representation learning capabilities and computing resource consumption is of great significance for the development of intelligent clinical diagnosis systems and computer vision.

Recently, structured state-space models (SSMs) \cite{37,38} inspired by classical state-space models have attracted great attention due to their computational efficiency and excellent performance in modeling long-range dependencies. Essentially, these models can be interpreted as a combination of recurrent neural networks (RNNs) and convolutional neural networks (CNNs) and can be computed very efficiently as either a recurrence or convolution, with linear or near-linear scaling in sequence length. Notably, Mamba \cite{39}, a state-of-the-art selective structured state space model, addresses the inherent limitations of previous SSMs, successfully demonstrates their efficiency and effectiveness in long sequence modeling, and achieves cutting-edge performance in continuous long sequence data analysis such as natural language processing and genomic analysis \cite{40}. From the perspective of internal mechanisms, Mamba incorporates time-varying parameters into the SSM and employs a novel hardware-aware algorithm to enable very efficient training and inference [41]. Crucially, it avoids the high quadratic computational complexity caused by the self-attention mechanism. Several recent studies have preliminarily explored the effectiveness of SSMs in various visual tasks, such as ImageNet classification\cite{41,42}, remote sensing image classification \cite{43}, image dehazing \cite{44}, point cloud analysis\cite{45}, and medical image segmentation\cite{40,46,47,48}. However, the potential of SSM in classifying medical images with different imaging modalities and morphological differences has not been fully studied. Given the impressive efficiency and powerful long-range dependency modeling capabilities of SSM, we attempted to employ SSM to replace Transformer to capture long-range dependencies in medical images efficiently. More precisely, we intuitively proposed a novel SSM-based generalized medical image classification model named MedMamba. Without too many fancy designs, MedMamba purely utilizes classic convolutional layers and SSM layers to extract local and global features of medical images from different modalities, respectively. By introducing the classic ideas of grouped convolution and channel-shuffle, MedMamba successfully provides fewer parameters and lower FLOPs for efficient medical artificial intelligence applications while maintaining excellent performance.

In summary, the contributions of this work can be summarized as follows: 1) To the best of our knowledge, this is the first research work that attempts to apply SSM to generalized medical image classification tasks and validates its effectiveness. 2) We proposed a novel hybrid basic block named SS-Conv-SSM. SS-Conv-SSM integrates channel-split, convolutional layers, SSM layers, and channel-shuffle to enables the model to extract features at all levels in medical images more efficiently. By repeatedly stacking SS-Conv-SSM block, the first vision Mamba for medical image classification, MedMamba, is introduced, including three variants (MedMamba-Tiny, MedMamba-Small, and MedMamba-Base). 3) 16 datasets with ten imaging modalities and 411,007 images and multiple performance metrics are used to evaluate MedMamba comprehensively. We fully observed the performance of SSM-based models on medical image datasets with different modalities and different data sizes. 4) We compared MedMamba with numerous state-of-the-art generic visual backbone networks. Experimental results show that MedMamba achieves competitive performance in various tasks. 5) We used t-SNE, Grad-CAM, and robustness analysis strategy to enhance the transparency and credibility of MedMamba, helping application developers and healthcare personnel to better understand and trust the model's decision-making process and facilitate the practical application of MedMamba. 6) Extensive experiments have demonstrated that the hybrid architecture based on convolution and SSM is more efficient in modeling medical images, laying a solid foundation for the design of future Mamba-based models for medical imaging.

\section{Related work}
\textbf{Convolution Neural Networks.} Convolutional Neural Network (CNN) is one of the most significant deep learning networks. The research on CNNs began in the 1980s and 1990s. Time Delay Neural Network \cite{49} and LeNet-5 \cite{50} were the earliest convolutional neural networks. After the 21st century, with the introduction of deep learning theory and improved numerical computing equipment, CNNs represented by AlexNet \cite{51}, Vgg \cite{52}, and ResNet \cite{53} have made significant breakthroughs in visual recognition tasks. The inspiration for CNN comes from the natural visual perception mechanism of living creatures, which enables them to automatically and effectively learn spatial hierarchical features from image data, making it suitable for various image recognition tasks \cite{54,55}. With the help of powerful computing devices (GPUs) and large-scale datasets, increasingly efficient CNNs have been proposed to enhance the practicality of AI systems in different scenarios \cite{56,57,58,59,60}. In the medical field, CNNs can learn and extract meaningful features from complex medical images and are widely used in skin disease diagnosis, cancer detection, histopathological analysis, infectious disease control, retinal disease identification, etc. \cite{61,62,63,64,65}. Utilizing well-designed CNNs for ImageNet and transfer learning strategies has become a common practice in medical image classification tasks. Today, CNN-based CAD can help doctors make more accurate diagnoses and improve patient treatment plans. With future technological advances and the continuously increasing medical data, CNNs will play an increasingly important role in improving the accuracy, efficiency, and automation of medical image analysis, opening up new possibilities for precision medicine and early diagnosis \cite{66}. 

\textbf{Vision Transformers.} Vision Transformer (ViT) is a deep learning model based on a self-attention mechanism, initially introduced by Google Research in 2020 for image recognition tasks \cite{67}. Unlike traditional CNNs, ViTs divide images into multiple patches and process them as sequences. This approach draws on the success of the Transformer model in the field of natural language processing. The proposal of ViTs marks a shift in deep learning models from relying on convolutional operations to utilizing self-attention mechanisms to process images. Early ViT-based models typically required large-scale datasets and appeared in simple configurations \cite{68}. Recent research on model design tends to incorporate inductive biases in visual perception into ViTs \cite{69,70,71,72,73}. In the medical field, ViTs are widely used in various tasks, such as medical image classification, medical image registration, and medical image segmentation, due to their ability to capture long-range dependencies and flexibly process images of different sizes \cite{15,34,36,74,75}. Since ViTs implement modality-agnostic modeling by treating an image as a sequence of patches without 2D inductive bias, it has become the preferred architecture for multimodal applications \cite{41}. At the same time, driven by the continuous growth of datasets, the increase in model size, and the advancement of model architecture, ViT-based foundation models provide unprecedented capabilities and are flexibly applied by researchers to multiple downstream tasks in the medical field \cite{76,77}. For example, researchers proposed MedSAM to segment everything in medical images \cite{78}. In contrast, RETGround has been proposed to identify and diagnose various diseases in the field of ophthalmology \cite{79}. In addition, UNI and PLIP are used to improve computational pathology \cite{80,81}. With more targeted improvements and optimizations, ViT is expected to play a greater role in medical image analysis.

\textbf{Visual State Space Models.} State Space Model (SSM) is a mathematical model used to describe and analyze the behavior of dynamic systems. Due to the need for a large amount of computation and memory, early SSMs were difficult to train and were not widely used in practice. Structured State Space Sequence Model (S4) \cite{37} employs a Normal Plus Low-Rank (NPLR) representation to efficiently compute the convolution kernel by leveraging the Woodbury identity for matrix inversion, addressing these limitations. As standalone sequence transformations, SSMs can be incorporated into end-to-end neural network architectures. Since then, various variants of SSMs have emerged \cite{82,83,84}. Unfortunately, the constant sequence transformation in SSM limits their context-based reasoning ability. Recent Mamba (Selective SSM) merges time-varying parameters into SSMs and formulates a hardware-aware algorithm for efficient training and inference \cite{39}, effectively enhancing the potential of SSMs. With the success of SSMs in long sequence modeling, researchers have begun to attempt to apply it to computer vision tasks, just like using the Transformer originally designed for NLP. ViS4mer \cite{85} and S4ND \cite{86} introduced the SSM blocks to facilitate modeling visual data across 1D, 2D, and 3D. VMamba \cite{42} proposed a pioneer vision backbone that integrates a cross-scan module with a mamba module to address the direction-sensitive problem arising from the differences between 1D sequences and 2D images. Similarly, Vim \cite{41} proposed bidirectional SSM for data-dependent global visual context modeling and incorporates position embeddings for location-aware visual understanding. Due to Mamba's impressive performance in visual tasks, researchers are actively applying Mamba to various fields. In the field of medical imaging, U-Mamba \cite{48} attempted to integrate Mamba layers into the encoder of nnUNet to enhance the modeling ability of CNN for long-range dependencies, and proposed a general-purpose network for both 3D and 2D biomedical image segmentation. Meanwhile, SegMamba \cite{47} combines the U-shape structure with Mamba to model the whole volume of global features at various scales. In addition, VM-UNet \cite{46} constructed the first medical image segmentation model based on a pure SSM-based model, establishing a new baseline for medical image segmentation. Swin-UMamba \cite{40} revealed how to enhance Mamba's performance in medical image segmentation using ImageNet-based training. The above studies demonstrate the promising results of SSM-based deep learning models in visual tasks.
\section{Methods}
\begin{figure}
    \centering
    \includegraphics[width=1\textwidth]{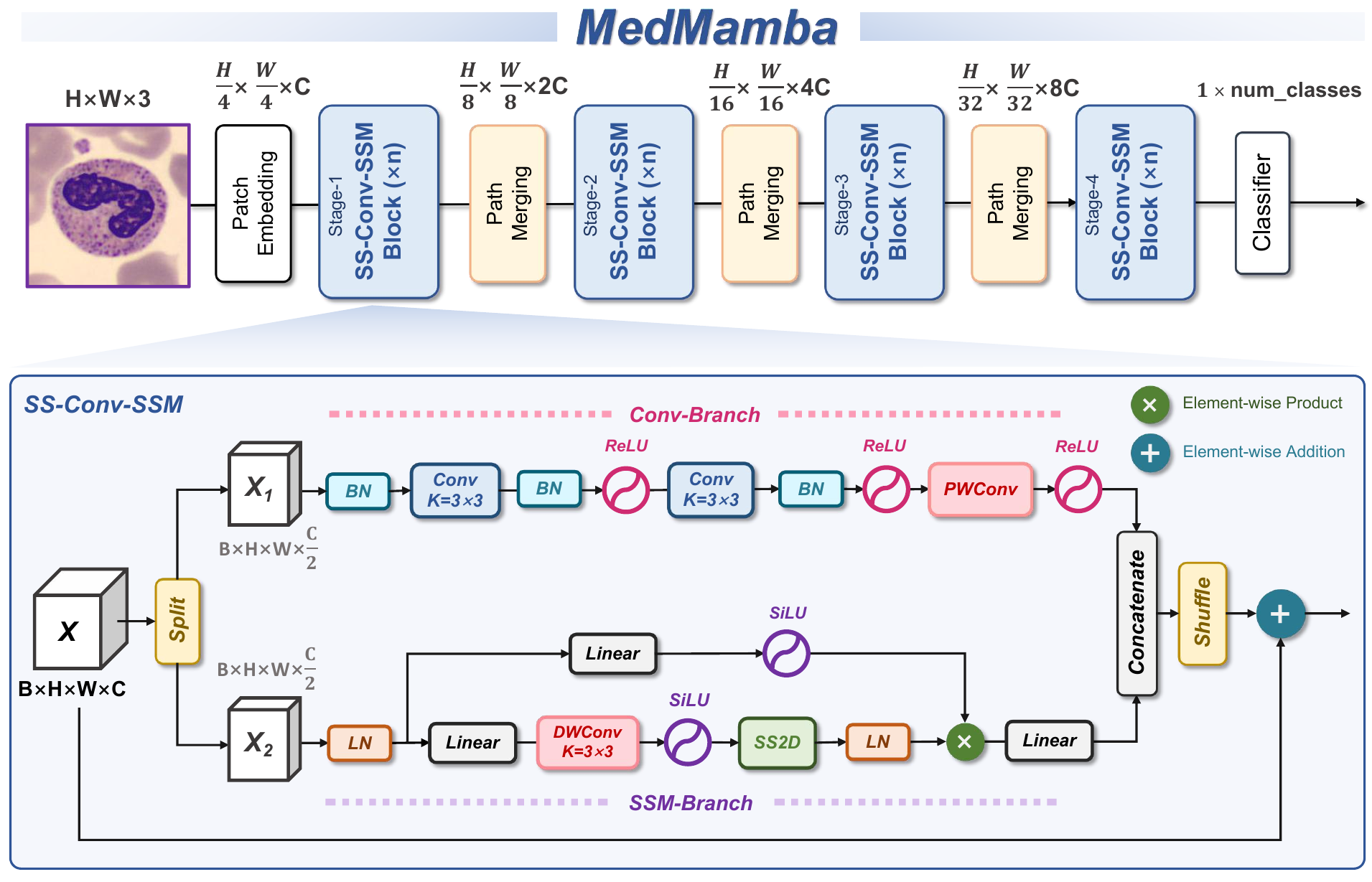}
    \caption{The overall architecture of the MedMamba. BN, LN, Linear, PWConv and DWConv represent Batch normalization, layer normalization, Linear layer, Point-wise convolution and Depth-wise convolution, respectively.}
    \label{fig1}
\end{figure}
In this section, we started by introducing the preliminary concepts related to MedMamba, i.e., the state space models and the discretization process. We then presented a comprehensive discussion for the overall architecture of MedMamba and the 2D-Selective-Scan mechanism. Finally, we provided detailed modelling process of the SS-Conv-SSM block which serves as the core element of MedMamba.
\subsection{Preliminaries}
The recent SSM-based models, i.e., Structured State Space Sequence Models (S4) and Mamba, rely on a classical continuous system that maps a one-dimensional input function or sequence, denoted as $x(t) \in \mathcal{R}$, through intermediate implicit states $h(t) \in \mathcal{R}^N$ to an output $y(t) \in \mathcal{R}$. The aforementioned process can be represented as a linear Ordinary Differential Equation (ODE) \cite{39,41,42,46}: 

\begin{equation}
\begin{aligned}
    &h^{\prime}(t) = \mathbf{A}h(t) + \mathbf{B}x(t) \\
    &y(t) = \mathbf{C}h(t)
\end{aligned}
\label{eq1}
\end{equation} where $\mathbf{A} \in \mathcal{R}^{N \times N}$ represents the state matrix, while $\mathbf{B} \in  \mathcal{R}^{N \times 1}$ and $\mathbf{C} \in  \mathcal{R}^{N \times 1}$ denote the projection parameters.

The S4 model and Mamba discretize the continuous system to make it more suitable for deep learning. Specifically, they introduce a timescale parameter $\mathbf{\Delta}$ to transform $\mathbf{A}$ and $\mathbf{B}$ into discrete parameters $\overline{\mathbf{A}}$ and $\overline{\mathbf{B}}$ using a fixed discretization rule. Typically, the zero-order hold (ZOH) is employed as the discretization rule and can be defined as follows:

\begin{equation}
\begin{aligned}
    &\overline{\mathbf{A}} = \textup{exp}(\mathbf{\Delta} \mathbf{A}) \\
    &\overline{\mathbf{B}} = (\mathbf{\Delta} \mathbf{A})^{-1}(\textup{exp}(\mathbf{\Delta} \mathbf{A}) - \mathbf{I})\cdot\mathbf{\Delta} \mathbf{B}
\end{aligned}
\label{eq2}
\end{equation}

After discretization, Equation \ref{eq1} utilizing a step size $\mathbf{\Delta}$ can be redefined as follows:

\begin{equation}
\begin{aligned}
    &h^{\prime}(t) = \overline{\mathbf{A}}h(t) + \overline{\mathbf{B}}x(t) \\
    &y(t) = \mathbf{C}h(t)
\end{aligned}
\label{eq:linear_recurrence}
\end{equation}
At the end of the process, the SSM model employ a global convolution to calculate the output: 
\begin{equation}
\begin{aligned}
    &\overline{K} = (\mathbf{C}\overline{\mathbf{B}}, \mathbf{C}\overline{\mathbf{AB}}, \ldots, \mathbf{C}\overline{\mathbf{A}}^{L-1}\overline{\mathbf{B}}) \\
    &y = x * \overline{\mathbf{K}}
\end{aligned}
\label{eq:global_convolution}
\end{equation}where $\overline{\mathbf{K}} \in \mathcal{R}^{L}$ represents a structured convolutional kernel, and $L$ denotes the length of the input sequence $x$.

\subsection{MedMamba}
\subsubsection{Overall architecture}
Figure \ref{fig1} shows the overall architecture of MedMamba. Specifically, MedMamba includes a patch embedding layer, stacked SS-Conv-SSM blocks, patch merging layers for down-sampling, and a feature classifier.

Similar to typical ViTs, MedMamba first partitions the input image 
$x \in {R^{H \times W \times 3}}$ into non-overlapping patches of size 4 × 4 using a patch embedding layer, subsequently mapping the channel dimensions of the image to C, but without further flattening the patches into a 1-D sequence. Patch embedding layer results in a feature map with $\frac{H}{4} \times \frac{W}{4} \times C$ without damaging the 2D structure of the input image.

Subsequently, MedMamba repeatedly stacks SS-Conv-SSM blocks to build Stage-1 that further process the feature map without changing the dimension size of it. To construct hierarchical representations, MedMamba utilizes a patch merging layer to down-sampling the feature map. Stage-2, Stage-3 and Stage-4 repeat the above process, and two patch merging layers are used for down-sampling the outputs of Stage-2 and Stage-3, resulting in two outputs with resolutions of $\frac{H}{{16}} \times \frac{W}{{16}} \times 4C$ and $\frac{H}{{32}} \times \frac{W}{{32}} \times 8C$ , respectively. At the end of the network, a classic classifier consisting of an adaptive global pooling layer and a linear layer is used to calculate the category of the input image. The entire process of MedMamba processing input is similar to the popular CNNs and ViTs. Besides, the default size for MedMamba's input is set as 224 x 224 x 3.

Like the configuration of most ViTs, three different scales of MedMamba were developed, i.e., MedMamba-Tiny, MedMamba-Small, and MedMamba-Base (referred to as MedMamba-T, MedMamba-S, and MedMamba-B, respectively). Detailed architectural specifications are outlined in Table \ref{tab1}. The FLOPs for all models are assessed using a 224 × 224 input size.

\newcommand{\blocka}[3]
{
\multirow{4}{*}{\(\left[\begin{array}{c}
\text{Channel Split}\\[-.1em]
\text{Conv-SSM}\\[-.1em]
\text{Channel Concat}\\[-.1em]
\text{Channel Shuffle}\\[-.1em]
\end{array}\right]\)$\times$#2}
}

\begin{table}[t]
\begin{center}
\resizebox{.95\textwidth}{!}
{
\begin{tabular}{c|c|c|c|c}
\toprule
\textbf{Layer name} & \textbf{Output Size} & \textbf{MedMamba-Tiny} & \textbf{MedMamba-Small} & \textbf{MedMamba-Base}  \\
\midrule
\multirow{2}{*}{Patch-E} & \multirow{2}{*}{56$\times$56} & conv 4$\times$4, 96, stride 4 & conv 4$\times$4, 96, stride 4 & conv 4$\times$4, 128, stride 4\\
\cline{3-5}
&  & \multicolumn{2}{c|}{Channel number $\rightarrow 96$} & Channel number $\rightarrow 128$ \\
\midrule
\multirow{6}{*}{Stage 1} & \multirow{6}{*}{56$\times$56} & &  \\ 
&  & \blocka{96}{2}{16}  & \blocka{96}{2}{16} & \blocka{128}{2}{16} \\
&  &  &  & \\
&  &  &  & \\
&  &  &  & \\
&  &  &  & \\
\midrule
Path-M& 28$\times$28 & \multicolumn{2}{c|}{Channel number $\rightarrow 192$} & Channel number $\rightarrow 256$ \\
\midrule

\multirow{6}{*}{Stage 2} & \multirow{6}{*}{28$\times$28} & &  \\ 
&  & \blocka{96}{2}{16}  & \blocka{96}{2}{16} & \blocka{128}{2}{16} \\
&  &  &  & \\
&  &  &  & \\
&  &  &  & \\
&  &  &  & \\
\midrule
Path-M& 14$\times$14 & \multicolumn{2}{c|}{Channel number $\rightarrow 384$} & Channel number $\rightarrow 512$ \\
\midrule

\multirow{6}{*}{Stage 3} & \multirow{6}{*}{14$\times$14} & &  \\ 
&  & \blocka{96}{4}{16}  & \blocka{96}{8}{16} & \blocka{128}{12}{16} \\
&  &  &  & \\
&  &  &  & \\
&  &  &  & \\
&  &  &  & \\
\midrule
Path-M& 7$\times$7 & \multicolumn{2}{c|}{Channel number $\rightarrow 768$} & Channel number $\rightarrow 1024$ \\
\midrule

\multirow{8}{*}{Stage 4} & \multirow{8}{*}{7$\times$7} & &  \\ 
&  & \blocka{768}{2}{16}  & \blocka{768}{2}{16} & \blocka{1024}{2}{16} \\
&  &  &  & \\
&  &  &  & \\
&  &  &  & \\
&  &  &  & \\
\midrule

Classifier & 1$\times$1  & \multicolumn{3}{c}{Adaptive global pooling, Linear (768/1024->num-classes), Softmax} \\
\midrule
\multicolumn{2}{c|}{Parameter Size(M)} & 15.2  & 23.5  & 48.1 \\
\midrule
\multicolumn{2}{c|}{FLOPs(G)} & 2.0$\times10^9$  & 3.5$\times10^9$  & 7.4$\times10^9$ \\
\bottomrule
\end{tabular}
}

\end{center}
%\vspace{-.5em}
\caption{
\textbf{Architectural overview of the MedMamba variants.} Patch-E and Patch-M represents Patch-Embedding and Patch-Merging, respectively. The results for parameter size and FLOPs are calculated when the number of classes is equal to 1000.}
\label{tab1}
\end{table}

\subsubsection{2D-selective-scan}
\begin{figure}
    \centering
    \includegraphics[width=1\textwidth]{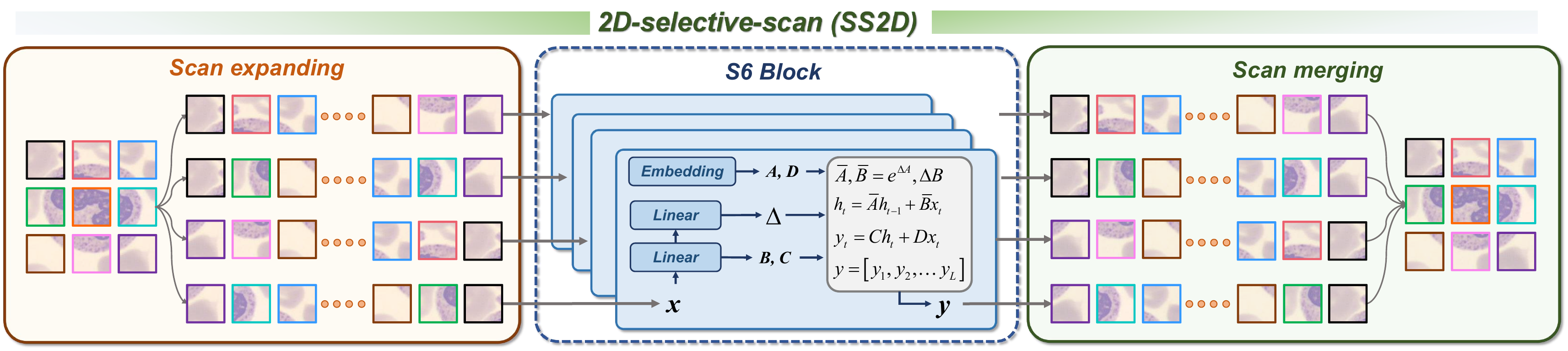}
    \caption{Visual display of the internal modeling process of SS2D.}
    \label{fig2}
\end{figure}

The 2D-selective-scan (SS2D) (Figure \ref{fig2}) proposed by VMamba\cite{42} is one of the core elements of MedMamba. SS2D inherits the selective scan space state sequence model (S6) designed for natural language processing tasks and successfully solves the "direction-sensitive" problem in S6. Specifically, to narrow the gap between 1-D array scanning and 2-D plain traversing, SS2D introduced a Cross-Scan Module (CSM), facilitating the extension of S6 to visual data without affecting the reception field. CSM adopts a four-way scanning strategy, i.e., from four corners all across the feature map to the opposite location to traverse the spatial domain of image feature maps, which ensures that each pixel in a feature map integrates information from all other locations in different directions, resulting in a global receptive field without increasing the linear computational complexity.

By integrating CSM, SS2D inherits the linear complexity of S6 while capturing long-range dependencies, which is essential for achieving accurate medical image classification. The SS2D contains three components consists: a scan expanding operation (CSM), an S6 block, and a scan merging operation. Figure2 visually presents the internal mechanism of SS2D. Concretely, the scan expanding operation first unfolds the input image along four different directions (top-left to bottom-right, bottom-right to top-left, top-right to bottom-left, and bottom-left to top-right) into sequences. An S6 block then processes all sequences to extract the features, ensuring that information from various directions is thoroughly scanned. Lastly, the four output features from the four directions are merged through scan merging to construct the final 2D feature map, resulting in a final output of the same size as the input. 

The S6 block originates from Mamba and introduces a selective mechanism based on S4 by adjusting the parameters of SSM according to input. This enables the model to distinguish and retain relevant information while filtering out irrelevant information. The pseudo-code for the S6 block is presented in Algorithm \ref{alg:s6}.

For ease of understanding, we denoted the input of SS2D, i.e., the feature map, as $I \in {R^{H \times W \times C}}$, where $I$[$h$][$w$] represents one token in the h-th row and w-th column of the feature map. Therefore, the scan expanding operation can be formalized as follows\cite{qin2024mambavc}: 
\begin{equation}
 {s_1}[i] = I\left[ {i\bmod W} \right]\left[ {\left\lfloor {i/W} \right\rfloor } \right] 
\end{equation}
\begin{equation}
{s_2}[i] = I\left[ {(N - i - 1)\bmod W} \right]\left[ {\left\lfloor {(N - i - 1)/W} \right\rfloor } \right]
\end{equation}
\begin{equation}
{s_3}[i] = I\left[ {\left\lfloor {i/H} \right\rfloor } \right]\left[ {\left\lfloor {i\bmod H} \right\rfloor } \right]
\end{equation}
\begin{equation}
{s_4}[i] = I\left[ {\left\lfloor {(N - i - 1)/H} \right\rfloor } \right]\left[ {(N - i - 1)\bmod H} \right]
\end{equation}
Where $N = H \times W$, $0 \le i < N$, ${s_1},{s_2},{s_3},{s_4} \in {R^{N \times C}}$ are the expanded and flattened token sequences. Next, we employed the S6 to selectively scan each token sequence, resulting in contextual token sequences ${s'_1},{s'_2},{s'_3},{s'_4} \in {R^{N \times C}}$. Immediately after, SS2D apply reversed operations to the contextual token sequences:
\begin{equation}
    {I'_1}[i][j] = {s'_1}[j \times W + i]
\end{equation}
\begin{equation}
{I'_2}[i][j] = {s'_2}[N - 1 - j \times W - i]
\end{equation}
\begin{equation}
{I'_3}[i][j] = {s'_3}[i \times H + j]
\end{equation}
\begin{equation}
{I'_4}[i][j] = {s'_4}[N - 1 - i \times H - j]
\end{equation}
Where ${I'_1},{I'_2},{I'_3},{I'_4} \in {R^{H \times W \times C}}$ denote the expanded and transformed feature map of I. In the end, SS2D applies the scan merging to obtain the output:
\begin{equation}
I' = {I'_1} + {I'_2} + {I'_3} + {I'_4}
\end{equation}

\begin{algorithm}[!t]
    \caption{Pseudo-code for S6 block in SS2D \cite{39,41,42,46}}
    \label{alg:s6}
    \renewcommand{\algorithmicensure}{\textbf{Output:}}
    \begin{algorithmic}[1]
        \renewcommand{\algorithmicrequire}{\textbf{Input:}}
        \REQUIRE $x$, the feature with shape [B, L, D] (batch size, token length, dimension)  %%input
        \renewcommand{\algorithmicrequire}{\textbf{Params:}}
        \REQUIRE $\mathbf{A}$, the nn.Parameter; $\mathbf{D}$, the nn.Parameter
        \renewcommand{\algorithmicrequire}{\textbf{Operator:}}
        \REQUIRE Linear(.), the linear projection layer
        \ENSURE $y$, the feature with shape [B, L, D]    %%output
        
        \STATE  $\mathbf{\Delta}, \mathbf{B}, \mathbf{C}$ = Linear($x$), Linear($x$), Linear($x$)
        \STATE  $\overline{\mathbf{A}} = \textup{exp}(\mathbf{\Delta} \mathbf{A})$
        \STATE  $\overline{\mathbf{B}} = (\mathbf{\Delta} \mathbf{A})^{-1}(\textup{exp}(\mathbf{\Delta} \mathbf{A}) - \mathbf{I})\cdot\mathbf{\Delta} \mathbf{B}$
        \STATE  $h_t = \overline{\mathbf{A}}h_{t-1} + \overline{\mathbf{B}}x_t$
        \STATE  $y_t = \mathbf{C}h_t + \mathbf{D}x_t$
        \STATE  $y = [y_1, y_2, \cdots, y_t, \cdots, y_L]$
        \RETURN $y$
    \end{algorithmic}
\end{algorithm}

\subsubsection{SS-Conv-SSM Block}

A grouped convolution uses a group of convolutions (multiple kernels per layer) to promote the model to learn various high- and low-level features. This concept was first introduced in AlexNet \cite{51} to distribute the model over multiple GPUs as an engineering compromise. In recent years, some studies have found that this module can be used to improve model performance while reducing model parameter size and model complexity. In order to enable MedMamba to model medical images more efficiently, we introduced this classic concept in the proposed SS-Conv-SSM. As a fundamental component of MedMamba, SS-Conv-SSM is a lightweight dual-branch block(Figure \ref{fig2}). It uses channel-split to partition the feature map into two groups and then uses Conv-Branch and SSM-Branch to extract global-local information from each group, respectively. Finally, SS-Conv-SSM employs channel-concatenation to restore the size of the channel dimension, while channel-shuffle is used to shuffle the feature map on the channel dimension to avoid information loss between channels caused by grouped convolution operations \cite{59,87}. We followed the settings of classic CNNs and ViTs and set the activation functions in Conv-Branch and SSM-Branch to ReLU \cite{88} and SilU \cite{89}, respectively.

We formalized the modeling process of SS-Conv-SSM for the feature maps. Given a module input $x \in {R^{H \times W \times C}}$ and a module output $y \in {R^{H \times W \times C}}$, we used $f$ to represent the channel-split, and then there is \[x \in {R^{H \times W \times C}}{x_{i = 1,2}} \in {R^{H \times W \times \frac{C}{2}}}\] Next, the ${f^{ - 1}}$ and $g$ are used to represent channel-concatenation and channel-shuffle respectively. To match the convolution operation, we utilized a permute operation to rearrange the original feature map. Based on the above, the modeling process of Conv-Branch can be defined as follows:
\[\overline {{x_1}}  \in {R^{\frac{C}{2} \times H \times W}} \leftarrow permute({x_1})\]
\[{x_1}^\prime  = BatchNor{m_1}(\overline {{x_1}} )\]
\[{x_1}^{\prime \prime } = ReLU(BatchNor{m_2}(Con{v_{3 \times 3}}({x_1}^\prime )))\]
\[{x_1}^{\prime \prime \prime } = ReLU(BatchNor{m_3}(Con{v_{3 \times 3}}({x_1}^{\prime \prime })))\]
\[\widehat {{x_1}} = ReLU(PWConv({x_1}^{\prime \prime \prime }))\]
\[\widetilde {{x_1}} \in {R^{H \times W \times \frac{C}{2}}} \leftarrow permute(\widehat {{x_1}})\]
Meanwhile, the modeling process of SSM-Branch can be defined as follows:
\[\overline {{x_2}}  = LayerNor{m_1}({x_2})\]
\[{x_2}^\prime  = SiLU(DWConv(Linear(\overline {{x_2}} )))\]
\[{x_2}^{\prime \prime } = LayerNor{m_2}(SS2D({x_2}^\prime ))\]
\[{x_2}^{\prime \prime \prime } = SiLU(Linear(\overline {{x_2}} ))\]
\[\widetilde {{x_2}} = Linear({x_2}^{\prime \prime } \otimes {x_2}^{\prime \prime \prime })\]
In summary, the output of SS-Conv-SSM be formulated as follows:
\[y = x \oplus g({f^{ - 1}}(\widetilde {{x_1}},\widetilde {{x_2}}))\]

\section{Experiments and results}
\subsection{Datasets}
\begin{figure}
    \centering
    \includegraphics[width=1\textwidth]{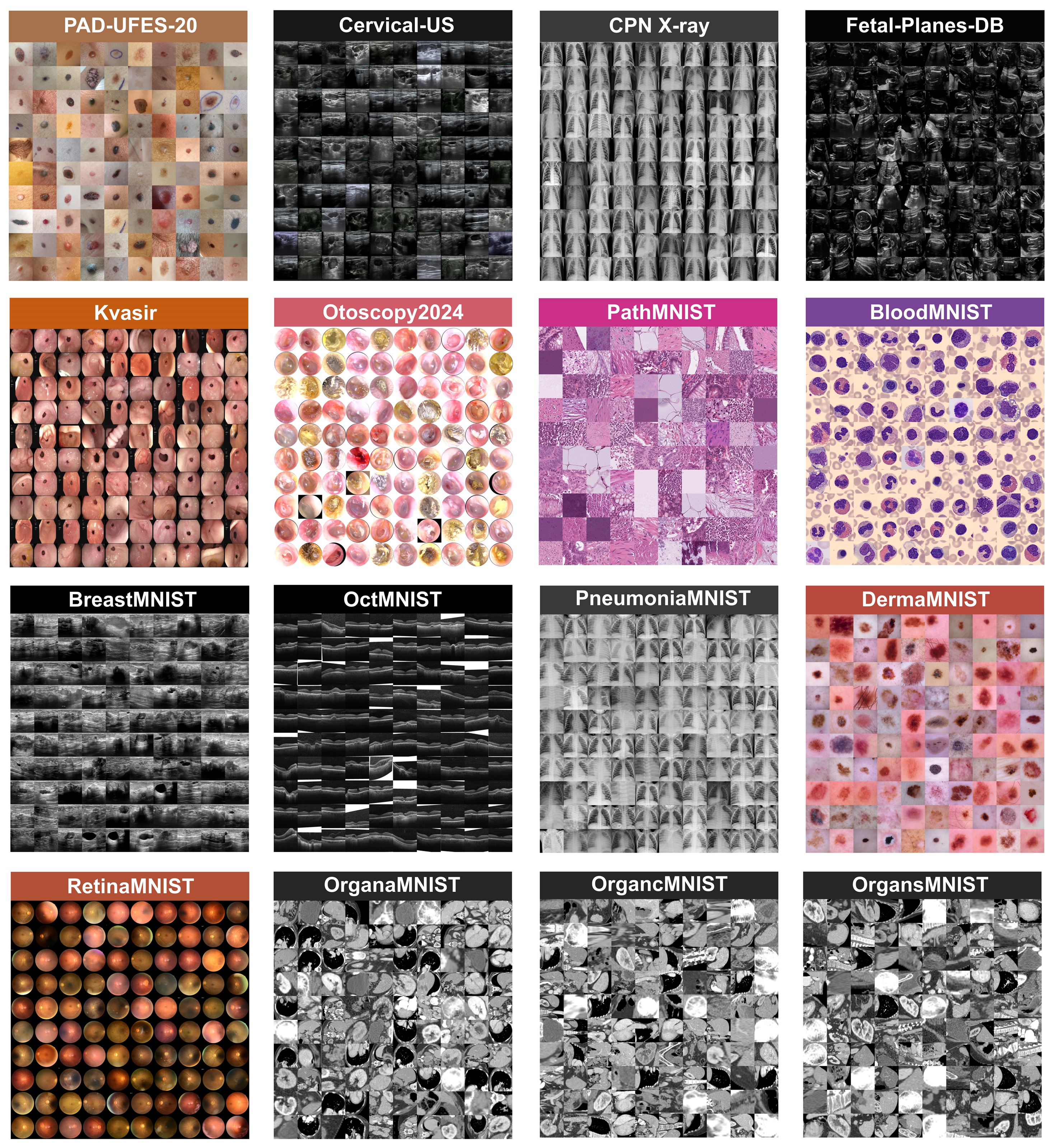}
    \caption{Typical samples of 16 different datasets with different imaging modalities.}
    \label{fig3}
\end{figure}

We adopted 16 medical image datasets (As shown in Figure \ref{fig3}), including two private datasets built by ourselves and 14 publicly available datasets to comprehensively evaluate the effectiveness and potential of MedMamba in medical image classification.

\textbf{PAD-UFES-20} \cite{90}. The PAD-UFES-20 is collected using different smartphone devices. The dataset consists of 2,298 samples of six different types of skin lesions. The skin lesions are: Basal Cell Carcinoma (BCC), Squamous Cell Carcinoma (SCC), Actinic Keratosis (ACK), Seborrheic Keratosis (SEK), Bowen’s disease (BOD), Melanoma (MEL), and Nevus (NEV).

\textbf{Cervical-US}. Cervical-US is a private ultrasound dataset built by ourselves, containing 3,392 Cervical lymph node lesion ultrasound images. Specifically, these images were obtained from 480 patients in the Ultrasound Department of the Second Affiliated Hospital of Guangzhou Medical University. The entire dataset is divided into four categories by clinical experts based on pathological biopsy results: normal lymph nodes (1,217 images), benign lymph nodes (607 images), malignant primary lymph nodes (236 images), and malignant metastatic lymph nodes (1,338 images). The institutional review board approved the use of the dataset.

\textbf{Fetal-Planes-DB} \cite{91}. A large dataset of routinely acquired maternal-fetal screening ultrasound images collected from two different hospitals by several operators and ultrasound machines. All images were manually labeled by an expert maternal fetal clinician. Images are divided into 6 classes: four of the most widely used fetal anatomical planes (Abdomen, Brain, Femur and Thorax), the mother’s cervix (widely used for prematurity screening) and a general category to include any other less common image plane. Fetal brain images are further categorized into the 3 most common fetal brain planes (Trans-thalamic, Trans-cerebellum, Trans-ventricular) to judge fine grain categorization performance.

\textbf{CPN X-ray} \cite{92,93}. CPN X-ray is a publicly available dataset containing 5,228 chest x-ray images. It helps the researcher and medical community to detect and classify COVID19 and Pneumonia from Chest X-Ray Images using Deep Learning. The entire dataset is divided into three categories: COVID-19, NORMAL, pneumonia.

\textbf{Kvasir} \cite{94}. The Kvasir dataset consists of images, annotated and verified by medical doctors (experienced endoscopists), including several classes showing anatomical landmarks, phatological findings or endoscopic procedures in the GI tract, i.e., hundreds of images for each class. The anatomical landmarks include Z-line, pylorus, cecum, etc., while the pathological finding includes esophagitis, polyps, ulcerative colitis, etc. In addition, Kvasir provides several sets of images related to removal of lesions, e.g., "dyed and lifted polyp", the "dyed resection margins", etc.

\textbf{Otoscopy2024} \cite{95}. Otoscopy2024 is a supplement to previous work. In previous work, we collected 20542 endoscopic images of ear infections. On this basis, we have added 2039 images from another medical institution. We name the new dataset as Otoscopy2024. Otoscopy2024 is a large dataset for ear disease classification, consisting of nine categories: Cholesteatoma of middle ear(548 images), Chronic suppurative otitis media(4,021 images), External auditory canal bleeding (451 images), Impacted cerumen (6,058 images), Normal eardrum (4,685 images), Otomycosis externa (2,507 images), Secretory otitis media (2,720 images), Tympanic membrane calcification (1,152 images), Acute otitis media (439 images).

\textbf{MedMNIST} \cite{96,97}. MedMNIST is a large-scale MNIST-like collection of standardized biomedical images, including 12 datasets for 2D and six datasets for 3D. All images have the corresponding classification labels, so that no background knowledge is required for users. The image size of this dataset includes the following sizes: 28x28 (2D), 64x64 (2D), 128x128 (2D), 224x224 (2D), 28x28x28 (3D) and 64x64x64 (3D). Covering primary data modalities in biomedical images, MedMNIST is designed to perform classification on lightweight 2D and 3D images with various data scales (from 100 to 100,000) and diverse tasks (binary/multi-class, ordinal regression and multi-label). Since MedMNIST is designed to be educational, standardized, diverse and lightweight, it could be used as a general classification benchmark in medical image analysis. In this work, we selected ten 2D medical image classification datasets: PathMNIST, DermaMNIST, OCTMNIST, PneumoniaMNIST, RetinaMNIST, BreastMNIST, BloodMNIST, OrganAMNIST, OrganCMNIST, and OrganSMNIST.

Table \ref{tab2} reports detailed information of each dataset, such as dataset size, image modality, and dataset splitting strategy. Here, the splitting strategy of MedMNIST is consistent with previous work\cite{15,97}.

\begin{table*}[!ht]
\scriptsize
\centering
\caption{The detailed descriptions for 16 datasets used in the work.}
\vspace{-0.7mm}
\label{tab2}
\resizebox{\textwidth}{!}{
\begin{tabular}{cccccccc}
\toprule
\textbf{Name}  & \textbf{Imaging Modality} & \textbf{Task (Classes)} &  \textbf{Dataset Size} & \textbf{Training/Validation/Test} & \textbf{Data Availability}\\
\midrule
PAD-UFES-20 &  Human Skin Smartphone Image & MC (6) & 2,298 & 1,384 / 227 / 687 &Public \\
cervical-US &  Cervical lymph Ultrasound & MC (4) & 3392 & 2,039 / 337 / 1,016 &Private \\
CPN X-ray &  Chet X-ray & MC (3) & 5,228 & 3,140 / 521 / 1,567 &Public \\
Fetal-Planes-DB &  Maternal-fetal Ultrasound & MC (6) & 1,2400 & 7,446 / 1,237 / 3,717 &Public \\
Kvasir &  Gastrointestinal Endoscope & MC (8) & 4,000 & 2,408 / 392 / 1,200 &Public \\
Otoscopy2024 &  Ear Endoscope & MC (9) & 24,233 & 1,4548 / 2,419 / 7,266 &Private \\
PathMNIST &  Colon Pathology & MC (9) & 107,180 & 89,996 / 10,004 / 7,180 &Public \\
OCTMNIST & Retinal OCT & MC (4) & 109,309 & 97,477 / 10,832 / 1,000  &Public\\
DermaMNIST & Dermatoscope & MC (7) & 10,015 & 7,007 / 1,003 / 2,005 &Public\\
PneumoniaMNIST & Chest X-Ray & BC (2) & 5,856 &  4,708 / 524 / 624 &Public\\
BreastMNIST & Breast Ultrasound & BC (2) & 780 & 546 / 78 / 156  &Public\\
RetinaMNIST & Fundus Camera & MC (5) & 1,600 & 1,080 / 120 /400 &Public \\
BloodMNIST & Blood Cell Microscope & MC (8) & 17,092 & 11,959 / 1,712 / 3,421  &Public\\
OrganAMNIST & Abdominal CT & MC (11) & 58,850 & 34,581 / 6,491 / 17,778  &Public\\
OrganCMNIST & Abdominal CT & MC (11) & 23,660 & 13,000 / 2,392 / 8,268 &Public \\
OrganSMNIST & Abdominal CT & MC (11) & 25,221 & 13,940 / 2,452 / 8,829 &Public\\
\bottomrule
\end{tabular}
}
\end{table*}

\subsection{Implementation Details}
We chose numerous well-designed generic vision networks that perform well on ImageNet to demonstrate the effectiveness and competitiveness of MedMamba in medical image classification. For a fair comparison, we resized all images to 224×224×3 before training each network. Subsequently, each dataset was normalized and standardized. 

During the training process of non-MedMNIST (the dataset that does not belong to MedMNIST), we employed the AdamW \cite{98} optimizer with a 0.0001 initial learning rate, B1 of 0.9, B2 of 0.999, and weight decay of 1e-4 and Cross-Entropy Loss to optimize the model parameters. To train the network, we utilized the PyTorch framework. We trained each model for 150 epochs and used a batch size of 64. We used an early-stop strategy to prevent model overfitting. Besides, we did not apply any data augmentation strategy and pre-training to demonstrate as much as possible that the results of all model metrics benefit from MedMamba's unique architecture. 

During the training process of MedMNIST, we followed the same training settings of MedMNISTv2 \cite{96} and MedViT \cite{15} without making any modifications to the original settings. Concretely, we trained all of the MedViT variants (MedMamba-T, MedMamba-S, and MedMamba-B) for 100 epochs and used a batch size of 128. We employed an AdamW optimizer with an initial learning rate of 0.001, the learning rate is decayed by a factor set of 0.1 in 50 and 75 epochs. In particular, to explore the impact of data augmentation strategies on the performance of MedMamba, we trained MedMamba-T exclusively using the Autoaugment \cite{99} strategy and denoted this new model as MedMamba-X. We then compared all MedMamba models with baseline methods and state-of-the-art methods on MedMNIST. All training processes were conducted on a computer with Ubuntu 22.04 operating system and four NVIDIA GeForce RTX 4090 GPUs.

\subsection{Evaluation Metrics}
We report Overall Accuracy, Precision, Sensitivity, Specificity, F1-score, and Area under the ROC Curve (AUC) as the standard evaluation metrics. \textbf{Overall Accuracy (OA)}: The proportion of all correct predictions (both true positives and true negatives) out of the total number of predictions. \textbf{Precision}: The proportion of true positive predictions out of all positive predictions made by the model.
\textbf{Sensitivity}: The proportion of true positive predictions made by the model out of all actual positive cases. \textbf{Specificity}: The proportion of true negative predictions out of all actual negative cases. \textbf{F1-score}: The harmonic mean of Precision and Sensitivity, balancing the contributions between two metrics. \textbf{AUC}: A metric that quantifies the overall performance of a binary classifier by measuring the area under the Receiver Operating Characteristic (ROC) curve, which plots the true positive rate against the false positive rate at various threshold settings. A higher AUC indicates better model performance in distinguishing between the positive and negative classes. 

The calculation formulas for Overall Accuracy, Precision, Sensitivity, Specificity, and F1-score are as follows:

\begin{equation}
Overall\ Accuracy = \frac{{TP + TN}}{{TP + TN + FP + FN}}    
\end{equation}

\begin{equation}
Precision = \frac{{TP}}{{TP + FP}}   
\end{equation}

\begin{equation}
Sensitivity = \frac{{TP}}{{TP + FN}}    
\end{equation}
\begin{equation}
Specificity = \frac{{TN}}{{TN + FP}}   
\end{equation}

\begin{equation}
F1-score = \frac{{2*Precision*Sensitivity}}{{Precision+Sensitivity}}   
\end{equation}
In addition, we also analyzed the model computation complexity (FLOPs) and the model parameter size.

\section{Results and discussions}
\subsection{The performance of MedMamba}

We divided reference models for non-MedMNIST into two categories: baseline models and non-baseline models. In this work, baseline models refer to models that have excellent performance in their respective architectures and are widely applied to downstream tasks. ConvNeXt, Swin Transformer (short for Swin) and VMamba were selected as baseline models. Non-baseline models refer to the latest state-of-the-art models (ViTs, CNNs and hybrid networks) with similar parameter sizes to MedMamba.

Table \ref{tab3} reports the metric results of MedMamba-T and reference models on PAD-UFES-20 and Cervical-US. On PAD-UFES-20, MedMamba-T with the lowest FLOPs achieves 58.8\% OA and 0.808 AUC. This result is extremely competitive among reference models. Compared with ConvNeXt-T, MedMamba-T has increased OA and AUC by 4.5\% and 0.048 respectively, while significantly reducing model complexity and parameter size. Notably, the AUC and OA of MedMamba-T surpass all non-baseline models. On the Cervical-US dataset, the performance of MedMamba-T is equally impressive. Compared with VMamba-T, Swin-T, and ConvNeXt-T, MedMamba-T achieves the best OA and presents an increase in OA by 1.8\%, 6.8\%, and 4.0\%, respectively. In terms of AUC, MedMamba's overall performance has surpassed all reference models except for VMamba-T.

\begin{table}[!t]
	\setlength\tabcolsep{3pt}
	\renewcommand\arraystretch{1.25}
	\caption{The performance comparison between MedMamba-t and reference models. Red font represents baseline models. Paras, P, Se, Sp, F1 means Parameter size, Precision, Sensitivity, Specificity and F1-score, respectively.}
	\begin{center}
 \resizebox{1\textwidth}{!}{%
		\begin{tabular}{c|c|cccccccc}
			\toprule\hline
\textbf{Dataset} &\textbf{Model} &\textbf{FLOPs(G)} &\textbf{Paras} & \textbf{P(\%)$\uparrow$}   & \textbf{Se(\%)$\uparrow$}   & \textbf{Sp(\%)$\uparrow$}   & \textbf{F1(\%)$\uparrow$}  & \textbf{OA(\%)$\uparrow$} & \textbf{AUC$\uparrow$} \\ \hline
   \multirow{11}{*}{PAD-UFES-20} 
                &\textbf{MedMamba-T}    &2.0  &14.5	&38.4	&36.9	&89.9	&35.8	&58.8	&0.808\\
&\textcolor{red}{VMamba-T}\cite{42}             &4.4	&22.1	&53.2	&40.6	&90.0	&41.6	&59.3	&0.804\\
&\textcolor{red}{Swin-T}\cite{69}               &4.5	&27.5	&38.2	&41.1	&90.6	&39.5	&60.5	&0.830\\
&\textcolor{red}{ConvNeXt-T}\cite{57}          &4.5	&27.8	&37.2	&33.6	&88.9	&33.7	&54.3	&0.760\\
&Repvgg-a1\cite{100}            &2.6	&12.8	&34.7	&37.7	&89.8	&35.9	&56.7	&0.803\\ 
&Mobilevitv2-200\cite{101} &5.6	&17.4	&33.9	&32.9	&88.0	&32.2	&49.9	&0.705 \\
&EdgeNext-base\cite{102}
&2.9	&17.9	&35.0	&36.4	&89.9	&34.6	&57.6	&0.802\\

&Nest-tiny\cite{103}
&5.8	&16.7	&49.9	&45.5	&91.3	&42.3	&63.5	&0.805\\

&Mobileone-s4\cite{104}
&3.0	&12.9	&35.9	&322	&87.9	&32.3	&49.3	&0.702\\

&Cait-xxs36\cite{105}
&3.8	&17.1	&37.1	&37.8	&90.0	&37.0	&58.6	&0.784\\

&DenseNet169\cite{106}
&3.4	&12.5	&44.5	&42.4	&90.1	&41.5	&58.0	&0.801\\
   
\hline\hline
\multirow{11}{*}{Cervical-US} 
 &\textbf{MedMamba-T}      &2.0	&14.5	&81.2	&76.2	&94.9	&78.0	&86.2	&0.952\\
&\textcolor{red}{VMamba-T}\cite{42}           & 4.4	 &22.1	&80.4	&74.2	&94.3	&76.4	&84.4	&0.953\\
&\textcolor{red}{Swin-T}\cite{69}             & 4.5	 &27.5	&72.0	&64.4	&92.2	&66.3	&79.4	&0.890\\
&\textcolor{red}{ConvNeXt-T}\cite{57}         &4.5	&27.8	&74.2	&68.8	&93.4	&70.6	&82.2	&0.931\\
&Repvgg-a1\cite{100}          &2.6	&12.8	&75.8	&73.6	&94.7	&74.6	&85.1	&0.937\\ 
&Mobilevitv2-200\cite{101}   &5.6	&17.4	&77.6	&67.4	&93.8	&68.1	&83.5	&0.904 \\
&EdgeNext-base\cite{102}
&2.9	&17.9	&71.1	&69.6	&92.6	&70.2	&79.3	&0.905\\

&Nest-tiny\cite{103}
&5.8	&16.7	&78.2	&69.8	&93.4	&72.3	&82.3	&0.926\\

&Mobileone-s4\cite{104}
&3.0	&12.9	&72.9	&69.3	&93.9	&70.3	&83.4	&0.918\\

&Cait-xxs36\cite{105}
&3.8	&17.1	&61.5	&57.4	&90.8	&58.0	&75.5	&0.860\\

&DenseNet169\cite{106}
&3.4	&12.5	&77.4	&78.8	&95.0	&78.0	&85.3	&0.947\\
\hline\bottomrule
		\end{tabular}}
		\label{tab3}
	\end{center}
\end{table}

Table \ref{tab4} reports the performance of MedMamba-S and reference models on CPN X-ray and Kvasir. On CPN X-ray, MedMamba-S with the lowest FLOPs achieves the best OA and AUC among all models. Compared with baseline models, the OA of MedMamba-S increases by 0.5\% (VMamba-S), 1.9\% (Swin-S) and 1.7\% (ConvNeXt-S), respectively, while maintaining the fewest parameters. Similarly, the performance of MedMamba-S is impressive on Kvasir. In terms of OA, MedMamba-T outperforms all reference models. Regarding AUC, MedMamba's result is only 0.1\% less than the top ranked Deit-small. 

\begin{table}[!t]
	\setlength\tabcolsep{3pt}
	\renewcommand\arraystretch{1.25}
	\caption{The performance comparison between MedMamba-S and reference models. Red font represents baseline models. Paras, P, Se, Sp, F1 means Parameter size, Precision, Sensitivity, Specificity and F1-score, respectively.}
	\begin{center}
 \resizebox{1\textwidth}{!}{%
		\begin{tabular}{c|c|cccccccc}
			\toprule\hline
\textbf{Dataset} &\textbf{Model} &\textbf{FLOPs(G)} &\textbf{Paras} & \textbf{P(\%)$\uparrow$}   & \textbf{Se(\%)$\uparrow$}   & \textbf{Sp(\%)$\uparrow$}   & \textbf{F1(\%)$\uparrow$}  & \textbf{OA(\%)$\uparrow$} & \textbf{AUC$\uparrow$} \\ \hline
   \multirow{12}{*}{CPN X-ray}

&\textbf{MedMamba-S}    &3.5	&22.8	&97.4	&97.4	&98.6	&97.4	&97.3	&0.997\\
&\textcolor{red}{VMamba-S}\cite{42} &9.0	&43.7	&96.8	&96.8	&98.3	&96.8	&96.8	&0.996 \\
&\textcolor{red}{Swin-S}\cite{69} &8.7	&48.8	&95.4	&95.5	&97.7	&95.4	&95.4	&0.993 \\
&\textcolor{red}{ConvNext-S}\cite{57} &8.7	&49.4	&95.7	&95.7	&97.8	&95.7	&95.6 	&0.994\\
&Convformer-s18\cite{107} &4.0	&24.7	&95.9	&95.8	&97.8	&95.8	&95.7	&0.992 \\ 
&TNT-s\cite{108} &5.2	&23.3	&93.4	&93.4	&96.6	&93.4	&93.2	&0.987 \\
&Caformer-s18\cite{107} &4.1	&24.3	&95.5	&95.5	&97.7	&95.5	&95.4	&0.992 \\
&PvtV2-b2\cite{109} &4.0	&24.8	&96.3	&96.2	&98.1	&96.2	&96.2	&0.994 \\
&Davit-tiny \cite{110}&4.5	&27.6	&95.1	&95.2	&97.5	&95.1	&95.1	&0.993 \\
&Deit-small\cite{111} &4.6	&21.7	&95.2	&95.1	&97.5	&95.1	&95.1	&0.990 \\

&EfficientNetV2-s\cite{112} &8.3	&20.2	&95.8	&95.7	&97.8	&95.7	&95.7	&0.993\\

&Coat-small\cite{113} &12.6	&21.4	&94.3	&94.2	&97.0	&94.2	&94.1	&0.987
\\
\hline\hline
\multirow{12}{*}{Kvasir} 
&\textbf{MedMamba-T}   &3.5	&22.8	&79.4	&79.3	&97.0	&79.2	&79.3	&0.976\\
&\textcolor{red}{VMamba-S}\cite{42}
&9.0	&43.7	&77.6	&77.3	&96.8	&77.1	&77.3	&0.970
\\
&\textcolor{red}{Swin-S}\cite{69}
&8.7	&48.8	&78.4	&78.0	&96.9	&77.3	&78.0	&0.973
\\
&\textcolor{red}{ConvNext-S}\cite{57}
&8.7	&49.4	&75.6	&74.8	&96.1	&74.8	&74.8	&0.969 \\
&Convformer-s18\cite{107}
&4.0	&24.7	&76.4	&75.8	&96.5	&75.6	&75.8	&0.970\\ 
&TNT-s\cite{108}
&5.2	&23.3	&76.5	&76.2	&96.6	&75.7	&76.2	&0.953
\\
&Caformer-s18\cite{107}
&4.1	&24.3	&73.6	&73.7	&96.2	&73.5	&73.7	&0.960
\\

&PvtV2-b2\cite{109}
&4.0	&24.8	&75.7	&75.6	&96.5	&75.3	&75.6	&0.958
\\

&Davit-tiny\cite{110}
&4.5	&27.6	&73.8	&73.6	&96.2	&73.0	&73.6	&0.966
\\

&Deit-small\cite{111}
&4.6	&21.7	&78.2	&78.1	&96.8	&77.9	&78.1	&0.977
\\

&EfficientNetV2-s\cite{112}
&8.3	&20.2	&78.7	&78.1	&96.8	&78.1	&78.2	&0.972\\

&Coat-small\cite{113}
&12.6	&21.4	&74.2	&73.5	&96.2	&73.1	&73.5	&0.969\\

\hline\bottomrule
		\end{tabular}}
		\label{tab4}
	\end{center}
\end{table}

Table \ref{tab5} reports the performance of MedMamba-B and reference models on Fetal-Planes-DB and Otoscopy2024. When the parameters are of similar size, MedMamba-B outperforms all reference models on both datasets. Specifically, MedMamba-B improves OA by 0.6\%, 5.2\%, and 5.3\% on Fetal-Planes-DB compared with VMamba-B, Swin-B, and ConvNeXt, respectively. Importantly, MedMamba-B reduces parameter size by 28.1M, 39.6M, and 40.5M, respectively, while having the lowest FLOPs, which is crucial for practical applications. Regarding the OA of Otoscopy2024, MedMamba-B outperforms all reference models except VMamba-B. Compared with Swin-B and ConvNeXt-B, Model C improves by 1.8% and 4.7% respectively.
\begin{table}[!t]
	\setlength\tabcolsep{3pt}
	\renewcommand\arraystretch{1.25}
	\caption{The performance comparison between MedMamba-B and reference models. Red font represents baseline models. Paras, P, Se, Sp, F1 means Parameter size, Precision, Sensitivity, Specificity and F1-score, respectively.}
	\begin{center}
 \resizebox{1\textwidth}{!}{%
		\begin{tabular}{c|c|cccccccc}
			\toprule\hline
\textbf{Dataset} &\textbf{Model} &\textbf{FLOPs(G)} &\textbf{Paras} & \textbf{P(\%)$\uparrow$}   & \textbf{Se(\%)$\uparrow$}   & \textbf{Sp(\%)$\uparrow$}   & \textbf{F1(\%)$\uparrow$}  & \textbf{OA(\%)$\uparrow$} & \textbf{AUC$\uparrow$} \\ \hline
   \multirow{13}{*}{Fetal-Planes-DB}

&\textbf{MedMamba-B}    &7.4	&47.1	&92.8	&93.8	&98.8	&93.3	&94.4	&0.993\\
&\textcolor{red}{VMamba-B}\cite{42}
&15.1	&75.2	&92.2	&93.4	&98.7	&92.7	&93.8	&0.994
\\
&\textcolor{red}{Swin-B}\cite{69}
&15.4	&86.7	&86.1	&84.9	&97.7	&85.4	&89.2	&0.982
\\
&\textcolor{red}{ConvNext-B}\cite{57}
&15.4	&87.6	&85.9	&85.2	&97.7	&85.5	&89.1	&0.982
\\
&Davit-small\cite{110}
&8.8	&48.9	&85.9	&84.8	&97.6	&85.3	&88.9	&0.977
\\ 
&Mvitv2-base\cite{114}
&10.0	&50.7	&89.9	&90.1	&98.3	&89.9	&91.9	&0.989
\\
&EfficientNet-b6\cite{115}
&19.0	&40.8	&91.2	&91.2	&98.4	&91.1	&92.8	&0.993
\\
&EfficientNetV2-b\cite{112}
&24.5	&52.9	&87.6	&89.1	&97.9	&88.3	&90.2	&0.983
\\
&FocalNet-s\cite{116}
&8.7	&49.1	&91.7	&90.9	&98.5	&91.2	&92.9	&0.989
\\
&Twins-SVT-base\cite{117}
&8.8	&48.9	&87.5	&88.4	&97.9	&88.0	&90.3	&0.987
\\

&Poolformer-m36\cite{118}
&8.8	&55.4	&82.7	&82.3	&87.4	&82.9	&87.67	&0.973
\\

&Xcit-s\cite{119}
&9.1	&47.3	&85.2	&86.1	&97.7	&85.5	&89.1	&0.981

\\
&GcVit-s\cite{120}
&8.4	&50.3	&84.5	&84.3	&97.5	&84.3	&88.4	&0.977
\\
\hline\hline
\multirow{13}{*}{Otoscopy2024} 
&\textbf{MedMamba-B}    &7.4	&47.1	&87.7	&83.7	&98.7	&85.4	&89.9	&0.988\\
&\textcolor{red}{VMamba-B}\cite{42}
&15.1	&75.2	&86.7	&92.9	&98.6	&84.6	&90.1	&0.988
\\
&\textcolor{red}{Swin-B}\cite{69}
&15.4	&86.7	&83.1	&80.8	&98.4	&81.7	&88.1	&0.987
\\
&\textcolor{red}{ConvNext-B}\cite{57}
&15.4	&87.6	&80.2	&75.8	&98.1	&77.6	&85.2	&0.980
\\
&Davit-small\cite{110}
&8.8	&48.9	&77.9	&73.5	&97.8	&75.4	&83.3	&0.971
\\ 
&Mvitv2-base\cite{114}
&10.0	&50.7	&83.5	&90.7	&98.4	&81.8	&88.1	&0.981\\
&EfficientNet-b6\cite{115}
&19.0	&40.8	&84.0	&81.4	&98.5	&82.5	&88.7	&0.989\\
&EfficientNetV2-b\cite{112}
&24.5	&52.9	&81.9	&78.5	&98.2	&79.8	&86.3	&0.981\\
&FocalNet-s\cite{116}
&8.7	&49.1	&82.9	&82.5	&98.4	&82.6	&87.9	&0.982\\
&Twins-SVT-base[]\cite{117}
&8.8	&48.9	&82.8	&78.3	&98.2	&80.3	&86.6	&0.979\\

&Poolformer-m36\cite{118}
&8.8	&55.4	&79.1	&73.0	&97.9	&75.4	&84.0	&0.969\\

&Xcit-s\cite{119}
&9.1	&47.3	&82.5	&80.3	&98.4	&81.4	&87.5	&0.982
\\
&GcVit-s\cite{120}
&8.4	&50.3	&84.5	&78.8	&98.3	&81.1	&87.5	&0.983\\

\hline\bottomrule
		\end{tabular}}
		\label{tab5}
	\end{center}
\end{table}

To more intuitively demonstrate the advantages of MedMamba, we presented the average OA and FLOPs of the MedMamba family and the baseline model on non-MedMNIST in Figure 4a. The experimental results show that the average OA of MedMamba-T, MedMamba-S, and MedMamba-B reach 84.0\%, 84.3\%, and 83.8\%, respectively. Regarding average OA, MedMamba variants improve by 0.8\%, 1.5\%, and 0.7\% compared with the three VMamba variants. Compared with the three Swin variants, MedMamba variants improve by 1.9\%, 1.3\%, and 1.8\%, respectively. Compared with the three ConvNeXt variants, MedMamba variants improve by 3.7\%, 4.2\%, and 4.0\%, respectively. As can be seen from Figure 4a, the closer the model is to the upper left corner of the figure, the better the overall performance of the model. Obviously, MedMamba has achieved a good trade-off between parameter size, FLOPs, and OA.

\begin{figure}
    \centering
    \includegraphics[width=1\textwidth]{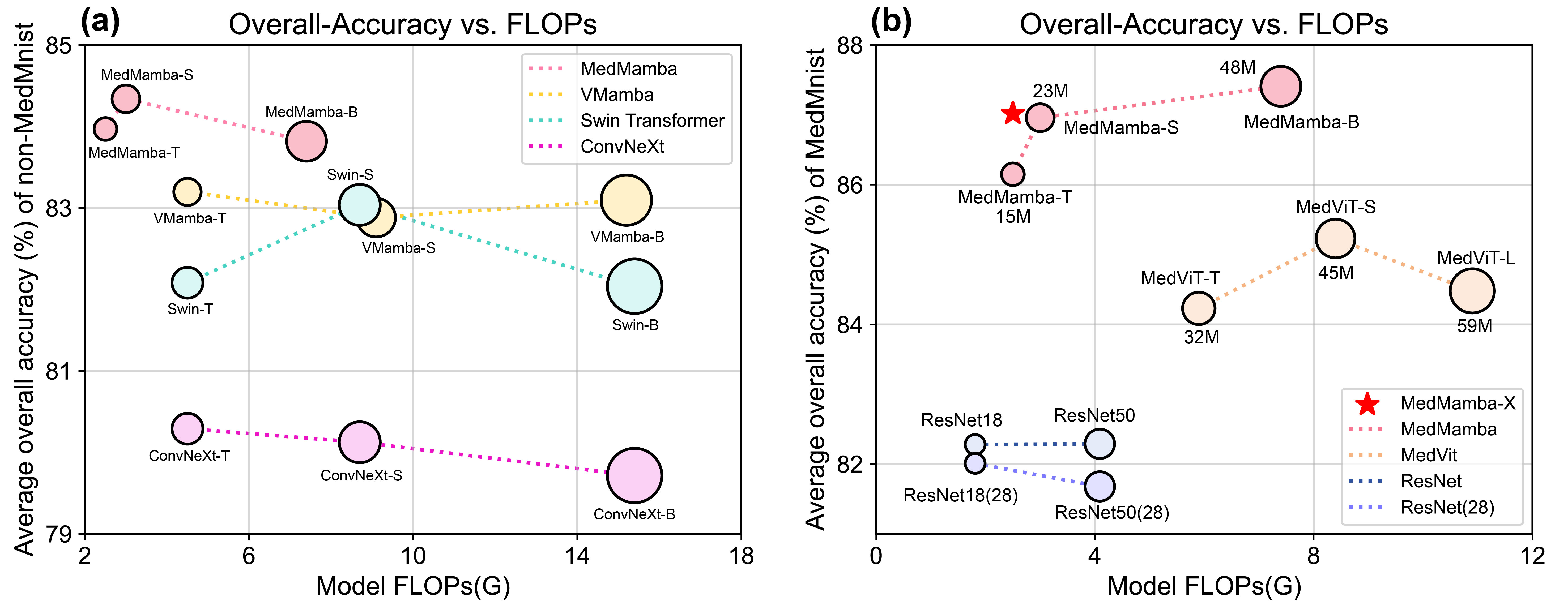}
    \caption{The performance comparison between MedMamba and reference models (Larger circles mean more parameters). a: The performance of MedMamba and baseline reference models on non-MedMNIST. b: The performance of MedMamba, MedViT and ResNet on MedMNIST.}
    \label{fig4}
\end{figure}

Table \ref{tab6} reports the performance comparison of MedMamba with previous state-of-the-art methods in terms of the AUC and ACC on each dataset of MedMNIST. Compared with the well-known ResNet18 and ResNet50, the three variants of MedMamba significantly improve the OA and AUC on each dataset. Taking PathMNIST as an example, the OA and AUC of MedMamba-S are 7.2\% and 1.0\% higher than ResNet50, respectively. Compared with the recent MedViT-S, the OA and AUC of MedMamba-S on OCTMNIST increase by 14.7\% and 3.6\%, respectively, which indicates that MedMamba may have a clear advantage in the OCT image-based classification tasks. Overall, our MedMamba effectively improves the performance of medical image classification tasks in the MedMNIST benchmark, especially PathMNIST, OCTMNIST, BloodMNIST, OrganaMNIST and OrgansMNIST. To more intuitively demonstrate the potential of MedMamba on MedMNIST, we presented each model's average OA and FLOPs in Figure 4b. Experimental results show that the average OA of MedMamba-T, MedMamba-S, and MedMamb-B reaches 86.2\%, 87.0\% and 87.4\% respectively. Compared with MedViT-T, MedViT-S, and MedViT-L, our MedMamba improves by 2.0\%, 1.7\% and 2.9\%, respectively. In particular, the FLOPs and parameter size of the MedMamba family are significantly lower than those of the MedViT family, which is helpful for practical applications and subsequent model improvements. As can be seen from Figure 4b, the closer the model is to the upper left corner of the figure, the better the overall performance of the model. Obviously, MedMamba has achieved a good trade-off between parameter size, FLOPs, and OA for MedMNIST. In summary, MedMamba performs well on most medical image classification tasks on MedMNIST, establishing a new baseline and achieving state-of-the-art results. Moreover, as shown in Table \ref{tab6} and Figure \ref{fig4}, when data augmentation is used, the average OA of MedMamba-X significantly surpasses the original MedMamba-T. Therefore, we recommend that future work adopt AutoAugment during model training.

\begin{table}[!ht]
    \centering
	\setlength\tabcolsep{3pt}
	\renewcommand\arraystretch{1.25}
	\caption{The performance of MedMamba-T on various datasets and comparison with reference models. (\textbf{Bold} font represents the best value. AOA represents the average overall accuracy of the model.)}
  \resizebox{1\textwidth}{!}{%
		\begin{tabular}{c||cc||cc||cc||cc||cc}
			\toprule \hline
			\multirow{2}{*}{\textbf{Methods}} &
			\multicolumn{2}{c||}{\textbf{PathMNIST}} &
			\multicolumn{2}{c||}{\textbf{DermaMNIST}} &
			\multicolumn{2}{c||}{\textbf{OCTMNIST}} &
			\multicolumn{2}{c||}{\textbf{PneumoniaMNIST}} &
			\multicolumn{2}{c}{\textbf{RetinaMNIST}} \\ \cline{2-11}
                 &AUC & OA & AUC & OA & AUC & OA & AUC & OA & AUC & OA \\ \midrule
			ResNet18 (28)\cite{53}&0.983	&0.907	&0.917	&0.735	&0.943	&0.743	&0.944	&0.854	&0.717	&0.524\\
			ResNet18 (224) \cite{53}&0.989	&0.909	&0.920	&0.754	&0.958	&0.763	&0.956	&0.864	&0.710	&0.493\\
			ResNet50 (28) \cite{53}&0.990	&0.911	&0.913	&0.735	&0.952	&0.762	&0.948	&0.854	&0.726	&0.528\\
                ResNet50 (224)\cite{53}& 0.989	&0.892	&0.912	&0.731	&0.958	&0.776	&0.962	&0.884	&0.716	&0.511\\
                Auto-sklearn\cite{121}  &0.934	&0.716	&0.902	&0.719	&0.887	&0.601	&0.942	&0.855	&0.690	&0.515 \\
                AutoKeras \cite{122}&0.959	&0.834	&0.915	&0749	&0.955	&0.763	&0.947	&0.878	&0.719	&0.503\\
                Google AutoML\cite{123}&0.944	&0.728	&0.914	&0.768	&0.963	&0.771	&0.991	&0.946	&0.750	&0.531\\
                MedVit-T \cite{15}&0.994	&0.938	&0.914	&0.768	&0.961	&0.767	&0.993	&0.949	&0.752	&0.534 \\
                MedVit-S  \cite{15}& 0.993	&0.942	&0.937	&0.780	&0.960	&0.782	&0.995	&0.961	&0.773	&0.561 \\
                MedVit-L \cite{15}&0.984	&0.933	&0.920	&0.773	&0.945	&0.761	&0.991	&0.921	&0.754	&0.552\\
   \textbf{MedMamba-T} &0.997	&0.953	&0.917	&0.779	&0.992	&0.918	&0.965	&0.899	&0.747	&0.543\\ 
   \textbf{MedMamba-S} &0.997	&0.955	&0.924	&0.758	&0.991	&0.929	&0.976	&0.936	&0.718	&0.545 \\ 
   \textbf{MedMamba-B} &0.999	&0.964	&0.925	&0.757	&0.996	&0.927	&0.973	&0.925	&0.715	&0.553 \\ 
   \textbf{MedMamba-X} &0.999	&0.962	&0.918	&0.751	&0.993	&0.928	&0.964	&0.910	&0.719	&0.570\\ 
   
   \midrule
   \multirow{2}{*}{\textbf{Methods}} &
			\multicolumn{2}{c||}{\textbf{BreastMNIST}} &
			\multicolumn{2}{c||}{\textbf{BloodMNIST}} &
			\multicolumn{2}{c||}{\textbf{OrganAMNIST}} &
			\multicolumn{2}{c||}{\textbf{OrganCMNIST}} &
			\multicolumn{2}{c}{\textbf{OrganSMNIST}} \\ \cline{2-11}
                 &AUC & OA & AUC & OA & AUC & OA & AUC & OA & AUC & OA \\ \midrule
                ResNet18 (28) \cite{53}&0.901	&0.863	&0.998	&0.958	&0.997	&0.935	&0.992	&0.900	&0.972	&0.782\\
			ResNet18 (224)\cite{53} &0.891	&0.833	&0.998	&0.963	&0.998	&0.951	&0.94	&0.920	&0.974	&0.778\\
			ResNet50 (28) \cite{53}&0.857	&0.812	&0.997	&0.956	&0.997	&0.935	&0.992	&0.905	&0.972	&0.770\\
                ResNet50 (224) \cite{53}&0.866	&0.842	&0.997	&0.950	&0.998	&0.947	&0.993	&0.911	&0.975	&0.785\\
                Auto-sklearn  \cite{121}&0.836	&0.803	&0.987	&0.878	&0.963	&0.762	&0.976	&0.829	&0.945	&0.672 \\
                AutoKeras \cite{122}&0.871	&0.831	&0.998	&0.961	&0.994	&0.905	&0.990	&0.879	&0.974	&0.813\\
                Google AutoML\cite{123}&0.919	&0.861	&0.998	&0.966	&0.990	&0.886	&0.988	&0.877	&0.964	&0.749\\
                MedVit-T\cite{15}&0.934	&0.896	&0.996	&0.950	&0.995	&0.931	&0.991	&0.901	&0.972	&0.789 \\
                MedVit-S\cite{15}&0.938	&0.897	&0.997	&0.951	&0.996	&0.928	&0.993	&0.916	&0.987	&0.805\\
                MedVit-L\cite{15}&0.929	&0.883	&0.996	&0.954	&0.997	&0.943	&0.994	&0.922	&0.973	&0.806\\
   \textbf{MedMamba-T} &0.825	&0.853	&0.999	&0.978	&0.998	&0.946	&0.997	&0.927	&0.982	&0.819\\ 
   \textbf{MedMamba-S} &0.806	&0.853	&0.999	&0.984	&0.999	&0.959	&0.997	&0.944	&0.984	&0.833 \\ 
   \textbf{MedMamba-B} &0.849	&0.891	&0.999	&0.983	&0.999	&0.964	&0.997	&0.943	&0.984	&0.834\\ 
   \textbf{MedMamba-X} &0.898	&0.853	&0.999	&0.984	&0.999	&0.967	&0.998	&0.947	&0.983	&0.830\\ 
            \hline\bottomrule
		\end{tabular}}
    \label{tab6}
\end{table}

\subsection{Visual Interpretation for MedMamba}
\begin{figure}
    \centering
    \includegraphics[width=1\textwidth]{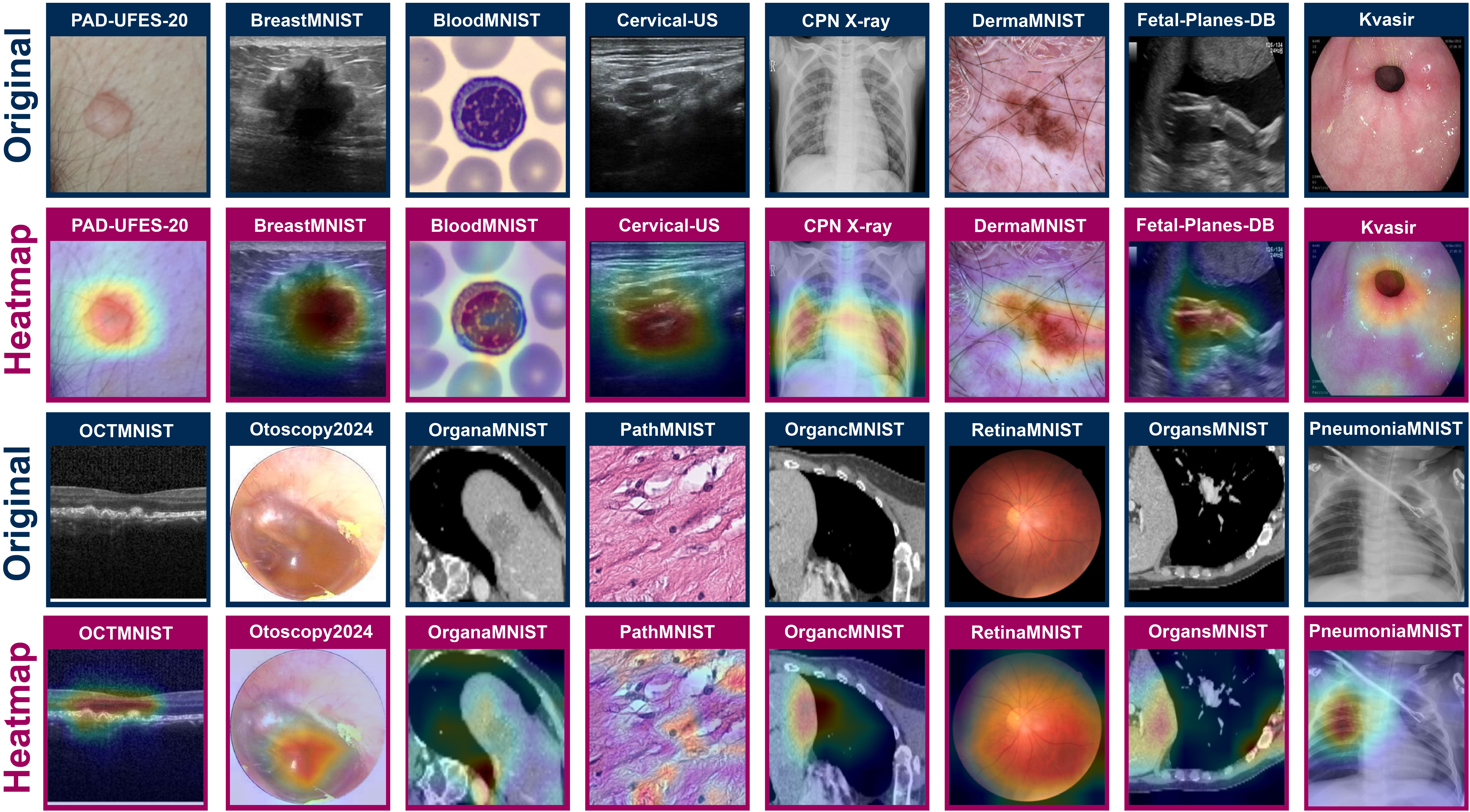}
    \caption{The visual heatmaps of MedMamba-S using Grad-CAM on 16 datasets.}
    \label{fig5}
\end{figure}

In order to enhance the transparency of MedMamba and enhance user trust in the model, we adopted Grad-CAM \cite{124} to visualize and examine MedMamba's internal decision-making mechanism. The Grad-CAM is presented with a rainbow colormap scale, where the red color stands for high relevance, yellow stands for medium relevance, and blue stands for low relevance. In Figure \ref{fig5}, we presented the heatmap results of MedMamba-S for typical images in each dataset. It is obvious that in most cases, our model can accurately focus on the lesion areas in each type of image, and rarely pay attention to the background parts that are not related to the prediction results.

\begin{figure}
    \centering
    \includegraphics[width=1\textwidth]{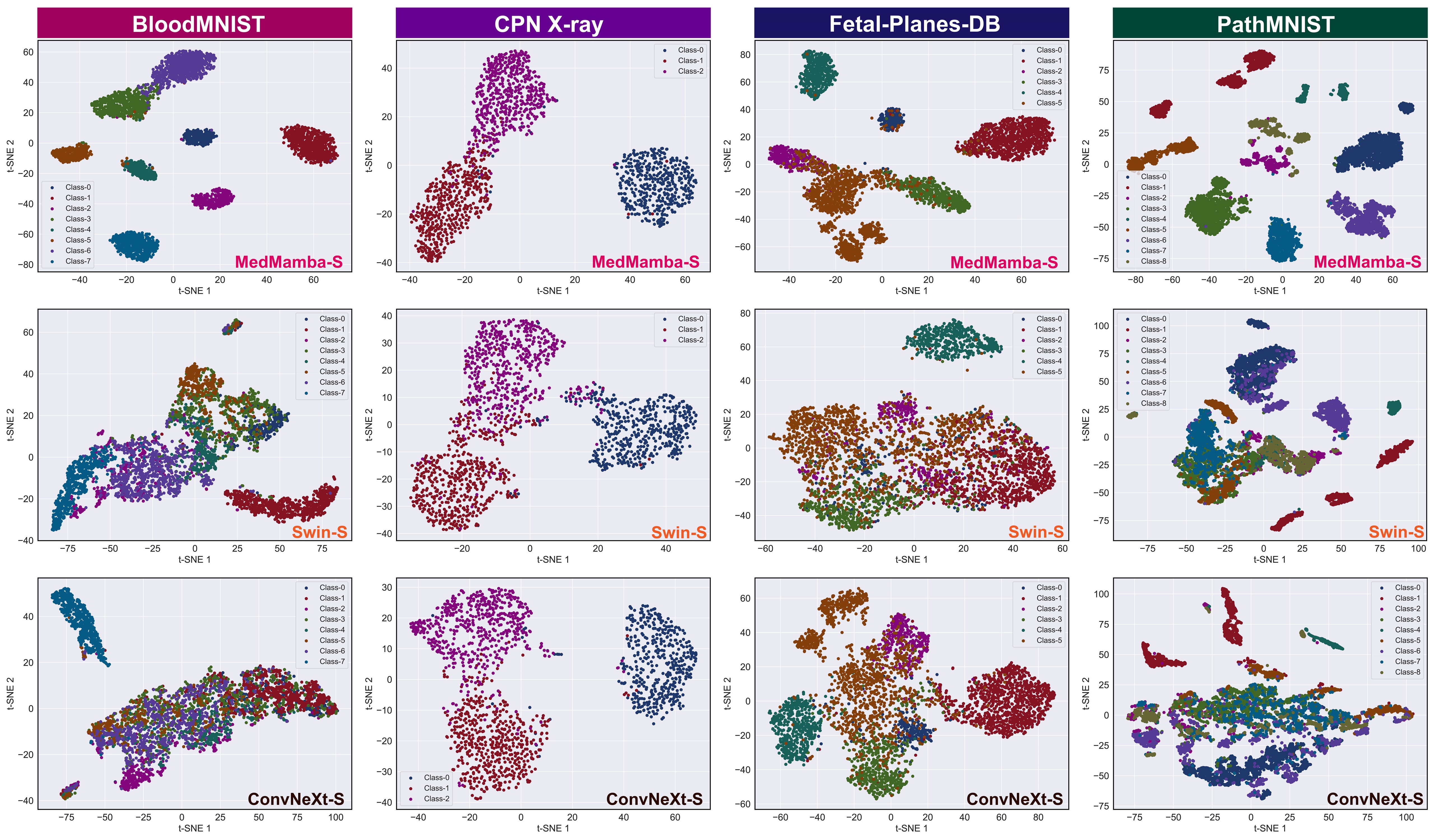}
    \caption{he t-SNE results of MedMamba-S and baseline models for CPN X-ray, Fetal-Planes-DB, BloodMNIST and PathMNIST.}
    \label{fig6}
\end{figure}

The t-SNE (t-distributed Stochastic Neighbor Embedding) is an unsupervised non-linear dimensionality reduction technique for data exploration and visualizing high-dimensional data. It is often used to visualize complex datasets into two and three dimensions, allowing researchers to understand more about underlying patterns and relationships in the data. We plotted the t-SNE results of MedMamba-S, Swin-S, ConvNeXt-S in Figure \ref{fig6}. The experimental results show that compared with the baseline models Swin-S and ConvNeXt-S, the features extracted by MedMamba are obviously more representative and discriminative. Specifically, in the t-SNE two-dimensional feature space constructed by MedMamba, sample points of the same category exhibit a significant clustering phenomenon, while sample points of different categories are relatively dispersed, reflecting MedMamba's effectiveness in capturing differences between categories. On the contrary, the baseline models Swin-S and ConvNeXt-S perform poorly in distinguishing samples from different categories, failing to effectively highlight the differences between categories, resulting in a more chaotic sample distribution. The above results indicate that MedMamba has better modeling ability for medical images.

\subsection{Robustness Analysis}

\begin{figure}
    \centering
    \includegraphics[width=1\textwidth]{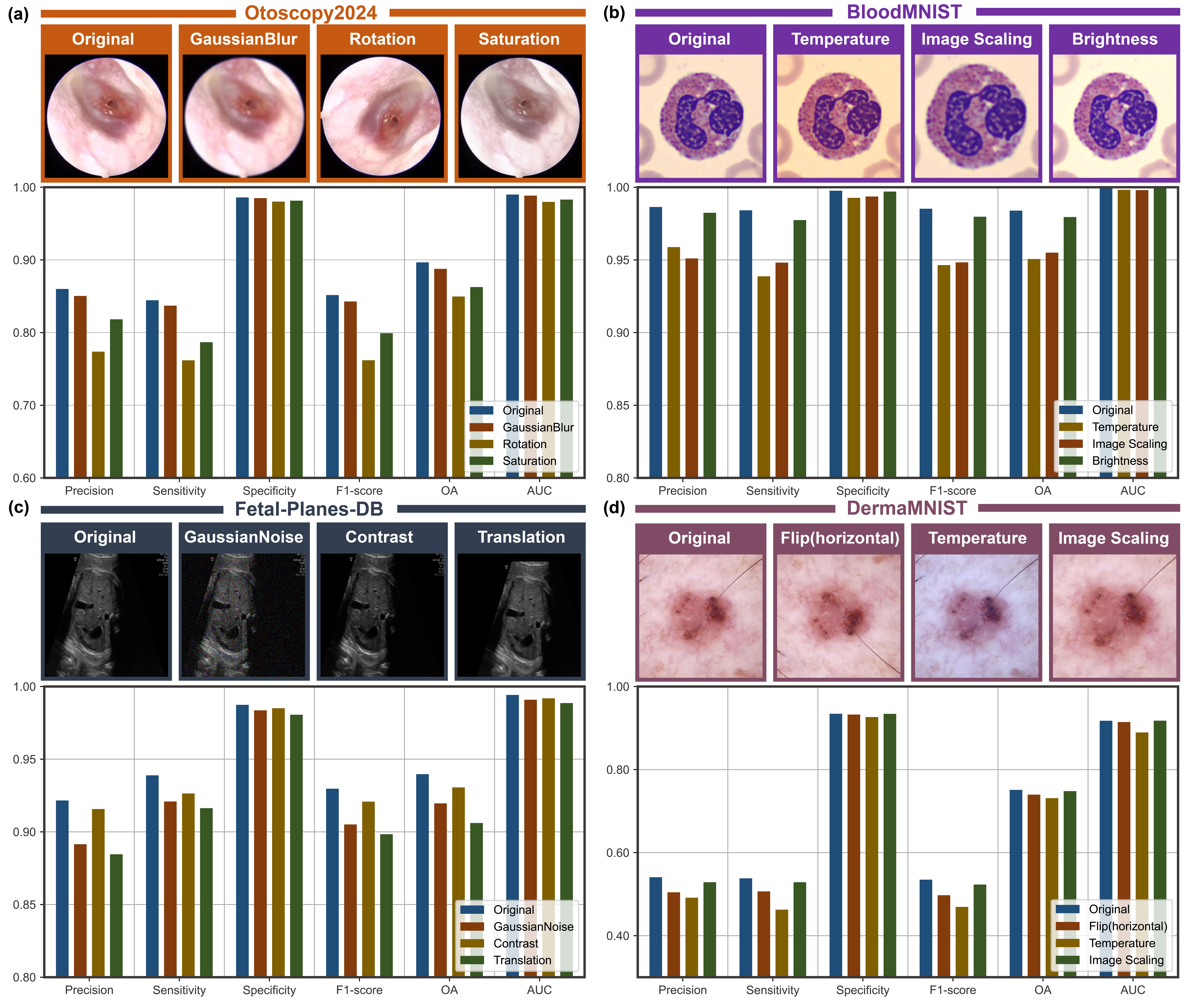}
    \caption{Performance comparison of MedMamba-X under different image perturbations. a: Robustness analysis results of Otoscopy2024. b: Robustness analysis results of BloodMNIST. c: Robustness analysis results of Fetal-Planes-DB. d: Robustness analysis results of DermMNIST.}
    \label{fig7}
\end{figure}

To facilitate the practical application and subsequent improvement of MedMamba, we referred to related work practices \cite{125,126,127} to analyze its robustness. Concretely, we artificially corrupted the original dataset to simulate various interferences that the model may encounter in practical clinical settings and observed the performance changes of the model under different perturbations. By testing MedMamba-X (As shown in Figure \ref{fig7}) on the Otoscopy2024, BloodMNIST, Fetal-Planes-DB, and DermMNIST datasets, we found the following phenomena: When corrupting the Otoscopy2024 dataset, MedMamba-X is particularly sensitive to the perturbation of image rotation, as shown by the downward trend in the results of all evaluation metrics. In contrast, for the BloodMNIST dataset, the color temperature change has the most significant impact on the performance of MedMamba-X. In particular, in terms of OA, the model's performance has decreased from 98.4\% to 95.1\%. On the Fetal-Planes-DB dataset, image translation has a more obvious impact on the performance of the model. Regarding the OA, the model's results decreased by 3.4\%. Similar to the findings on BloodMNIST, when perturbations based on color temperature changes are introduced, the performance of the model on all evaluation metrics decline. Overall, however, MedMamba maintains relatively stable performance in most cases of image perturbation.

\subsection{Ablation Study}

We conducted various ablation experiments to investigate the effectiveness of the critical components of our architecture. Table \ref{tab7} reports in detail the performance changes of MedMamba-X with different components. Specifically, when only SSM is used (denoted as Base Model-I), all metric results of the model are at the lowest level. Notably, when we integrated the Conv-branch on its basis (denoted as Base Model-II) and directly used element-wise addition to fuse the results of the SSM-branch and the Conv-branch, the model's performance was significantly improved. Taking PathMNIST and Otoscopy2024 as examples, the OA of Base Model-II is improved by 1.5\% and 1.3\%, respectively compared with Base Model-I. Unfortunately, this simple strategy will lead to a sharp increase in model complexity and parameter size, i.e., the parameter size and FLOPs of the model increased by 35.4M and 5.4G, respectively. This problem has been effectively alleviated by introducing the idea of grouped convolution, i.e., the channel-split operation. Furthermore, the channel-shuffle operation further strengthens the interaction of channel information between different groups and improves the performance of the model without increasing FLOPs and parameter size. Such improvements demonstrate that while optimizing model performance, we have also effectively controlled the complexity and parameter size of the model, achieving a balance between performance and efficiency.

\begin{table}[!t]
\setlength{\tabcolsep}{10pt} % 设置列间距为10磅  

	\caption{The performance of MedMamba-T with different components for various datasets. Paras represents parameter size. ‘SP’ and ‘SF’ represents Channel-Split and Channel-Shuffle, respectively.}
	\begin{center}
 \resizebox{0.8\textwidth}{!}{%
		\begin{tabular}{c|cccc|cccc}
		\toprule\hline
                \multirow{2}{*}{\textbf{Dataset}} &
			\multicolumn{4}{c|}{\textbf{Components}} &
			\multicolumn{4}{c}{\textbf{Evaluation metrics}}\\
  &\textbf{SSM} &\textbf{Conv} &\textbf{SP} &\textbf{SF} &\textbf{FLOPs(G)$\downarrow$} &\textbf{Paras(M)$\downarrow$} &\textbf{OA(\%)$\uparrow$} &\textbf{AUC$\uparrow$}

\\ \hline
\multirow{4}{*}{Cervical-US} 
&\textcolor[RGB]{0,153,76}{\checkmark} &\textcolor{red}{\ding{55}}	&\textcolor{red}{\ding{55}} &\textcolor{red}{\ding{55}} &2.0	&16.2	&83.9	&0.939\\
&\textcolor[RGB]{0,153,76}{\checkmark} &\textcolor[RGB]{0,153,76}{\checkmark}	&\textcolor{red}{\ding{55}} &\textcolor{red}{\ding{55}} &7.4	&51.6	&85.1	&0.937\\
&\textcolor[RGB]{0,153,76}{\checkmark} &\textcolor[RGB]{0,153,76}{\checkmark}	&\textcolor[RGB]{0,153,76}{\checkmark} &\textcolor{red}{\ding{55}} &2.0	&14.5	&85.4	&0.941\\
&\textcolor[RGB]{0,153,76}{\checkmark} &\textcolor[RGB]{0,153,76}{\checkmark}	&\textcolor[RGB]{0,153,76}{\checkmark}&\textcolor[RGB]{0,153,76}{\checkmark} &2.0	&14.5	&85.6	&0.953 \\ \hline
\multirow{4}{*}{PathMNIST} 
&\textcolor[RGB]{0,153,76}{\checkmark} &\textcolor{red}{\ding{55}}	&\textcolor{red}{\ding{55}} &\textcolor{red}{\ding{55}}  &2.0	&16.2	&94.4	&0.993\\
&\textcolor[RGB]{0,153,76}{\checkmark} &\textcolor[RGB]{0,153,76}{\checkmark}	&\textcolor{red}{\ding{55}} &\textcolor{red}{\ding{55}} &7.4	&51.6	&95.9	&0.997\\
&\textcolor[RGB]{0,153,76}{\checkmark} &\textcolor[RGB]{0,153,76}{\checkmark}	&\textcolor[RGB]{0,153,76}{\checkmark} &\textcolor{red}{\ding{55}} &2.0	&14.5	&95.0	&0.994\\
&\textcolor[RGB]{0,153,76}{\checkmark} &\textcolor[RGB]{0,153,76}{\checkmark}	&\textcolor[RGB]{0,153,76}{\checkmark}&\textcolor[RGB]{0,153,76}{\checkmark} &2.0	&14.5	&95.3	&0.997 \\ \hline
\multirow{4}{*}{CPN X-ray} 
&\textcolor[RGB]{0,153,76}{\checkmark} &\textcolor{red}{\ding{55}}	&\textcolor{red}{\ding{55}} &\textcolor{red}{\ding{55}}  &2.0	&16.2	&96.3	&0.993\\
&\textcolor[RGB]{0,153,76}{\checkmark} &\textcolor[RGB]{0,153,76}{\checkmark}	&\textcolor{red}{\ding{55}} &\textcolor{red}{\ding{55}} &7.4	&51.6	&97.4	&0.997\\
&\textcolor[RGB]{0,153,76}{\checkmark} &\textcolor[RGB]{0,153,76}{\checkmark}	&\textcolor[RGB]{0,153,76}{\checkmark} &\textcolor{red}{\ding{55}} &2.0	&14.5	&96.8	&0.995\\
&\textcolor[RGB]{0,153,76}{\checkmark} &\textcolor[RGB]{0,153,76}{\checkmark}	&\textcolor[RGB]{0,153,76}{\checkmark}&\textcolor[RGB]{0,153,76}{\checkmark} &2.0	&14.5	&97.1	&0.995 \\ \hline
\multirow{4}{*}{OCTMNIST} 
&\textcolor[RGB]{0,153,76}{\checkmark} &\textcolor{red}{\ding{55}}	&\textcolor{red}{\ding{55}} &\textcolor{red}{\ding{55}} &2.0	&16.2	&91.4	&0.986\\
&\textcolor[RGB]{0,153,76}{\checkmark} &\textcolor[RGB]{0,153,76}{\checkmark}	&\textcolor{red}{\ding{55}} &\textcolor{red}{\ding{55}} &7.4	&51.6	&92.3	&0.995\\
&\textcolor[RGB]{0,153,76}{\checkmark} &\textcolor[RGB]{0,153,76}{\checkmark}	&\textcolor[RGB]{0,153,76}{\checkmark} &\textcolor{red}{\ding{55}} &2.0	&14.5	&91.5	&0.993\\
&\textcolor[RGB]{0,153,76}{\checkmark} &\textcolor[RGB]{0,153,76}{\checkmark}	&\textcolor[RGB]{0,153,76}{\checkmark}&\textcolor[RGB]{0,153,76}{\checkmark} &2.0	&14.5	&91.8	&0.992 \\ \hline
\multirow{4}{*}{Otoscopy2024} 
&\textcolor[RGB]{0,153,76}{\checkmark} &\textcolor{red}{\ding{55}}	&\textcolor{red}{\ding{55}} &\textcolor{red}{\ding{55}}  &2.0	&16.2	&88.4	&0.983\\
&\textcolor[RGB]{0,153,76}{\checkmark} &\textcolor[RGB]{0,153,76}{\checkmark}	&\textcolor{red}{\ding{55}} &\textcolor{red}{\ding{55}}&7.4	&51.6	&89.7	&0.990\\
&\textcolor[RGB]{0,153,76}{\checkmark} &\textcolor[RGB]{0,153,76}{\checkmark}	&\textcolor[RGB]{0,153,76}{\checkmark} &\textcolor{red}{\ding{55}} &2.0	&14.5	&89.1	&0.985\\
&\textcolor[RGB]{0,153,76}{\checkmark} &\textcolor[RGB]{0,153,76}{\checkmark}	&\textcolor[RGB]{0,153,76}{\checkmark}&\textcolor[RGB]{0,153,76}{\checkmark}&2.0	&14.5	&89.5	&0.989\\
\hline\bottomrule
		\end{tabular}}
		\label{tab7}
	\end{center}
\end{table}

\section{Conclusion}
In summary, we proposed the MedMamba family in this work, which is the first vision mamba for general medical image classification. Although customized architectures designed for specific medical image datasets are superior in accuracy and efficiency, our MedMamba is able to extract local feature representations and long-range dependencies for performing generalized medical image classification tasks by combining classic convolutional layers and SSM modules. In particular, by introducing the grouped convolution strategy inside it, MedMamba achieves a good trade-off between efficient modeling and computational resource consumption. Extensive experiments show that our MedMamba achieves extremely competitive performance on 16 datasets containing ten image modalities and 411,007 images compared with state-of-the-art methods. We hope that this work can provide inspiration for future work in Mamba-based and model architecture design and intelligent medical applications. 

In addition, we summarize our future work into the following points: 1) We will further explore and test the potential of MedMamba on medical datasets obtained from other imaging modalities, and optimize the internal architecture of MedMamba. 2) We will further use explainable artificial intelligence to analyze MedMamba's decision-making mechanism. In addition, the impact of MedMamba's inference speed and model parameter size on practical applications should also be studied. 3) Due to the advantages of SSM in modeling long-range dependencies, further research on the performance of MedMamba in some high-resolution medical images (such as pathological images) may be beneficial. 4) We will attempt to use the backbone of MedMamba as an encoder or decoder to explore its application potential in more advanced medical image tasks, such as medical image segmentation, medical object detection, medical image registration, and medical image reconstruction.

\section*{Data Availability}
The codes and pre-trained weights of MedMamba family and access links to public datasets are available at https://github.com/YubiaoYue/MedMamba. The Cervical-US and Otoscopy2024 that support the findings of this study are available on request from the corresponding author upon reasonable request.

\section*{Acknowledgments}
This work is supported NSF of Guangdong Province (No.2022A1515011044 and No. 2023A1515010885), the project of promoting research capabilities for key construtted disciplines in Guangdong Province (No. 2021ZDJS028), Guangzhou Science and Technology Plan Project (No. 202201011696), the scientific research capacity improvement project of the doctoral program construction unit of Guangdong Polytechnic Normal University in 2022 (No. 22GPNUZDJS31). In addition, we would like to express our gratitude to VMamba, VM-UNet, and Swin-UMamba for their open source.
%Bibliography
\bibliographystyle{unsrt}  
\bibliography{references}

\end{document}